\documentclass[aps,prx,twocolumn,10pt,longbibliography,floatfix,superscriptaddress,eqsecnum]{revtex4-2}
\usepackage{amsmath,amssymb,amsfonts,float,epsfig,epstopdf,color,verbatim,tabularx,multirow,appendix}
\usepackage{bm}

\usepackage{graphicx}% Include figure files
\usepackage{xcolor}
\usepackage{dcolumn}% Align table columns on decimal point
\usepackage{bm}% bold math
\usepackage{amsmath}
\usepackage{xspace}
\usepackage{time}
\usepackage{comment}
\usepackage{textgreek}
\usepackage{color}
\definecolor{LinkColor}{rgb}{0.256,0.439,0.588}
\usepackage{hyperref}
\hypersetup{
colorlinks=true,
citecolor=LinkColor,
linkcolor=LinkColor,
urlcolor=LinkColor
}

\usepackage{amsmath}
\usepackage{mathtools}
\usepackage{amsfonts}
\usepackage{bm}
\usepackage{amssymb}

\usepackage{newtxmath}
\usepackage[thinlines]{easytable}
\usepackage{booktabs}
\usepackage{color}
\usepackage{xcolor}
\definecolor{darkblue}{rgb}{0.0, 0.0, 0.45}
\definecolor{darkred}{rgb}{0.64, 0.0, 0.0}
\definecolor{blue}{rgb}{0.2, 0.3, 0.85}
\definecolor{red}{rgb}{0.2, 0.3, 0.85}
\definecolor{darkgreen}{rgb}{0.0, 0.5, 0.0}
\usepackage{multibib}
\usepackage{todonotes}
\usepackage{color}
\usepackage{xcolor}
\definecolor{blue}{rgb}{0.25, 0.3, 0.95}
\definecolor{red}{rgb}{1,0,0}
\definecolor{darkgreen}{rgb}{0.0, 0.5, 0.0}

\usepackage{braket}
\usepackage{graphicx} %Loading the package

\newcommand{\dagga}{\vphantom{\dagger}}
\usepackage{tikz}
\usepackage{textcomp}
\def\be{\begin{equation}}
\def\ee{\end{equation}}
\def\bea{\begin{eqnarray}}
\def\eea{\end{eqnarray}}

\begin{document}

%\title{Strong Zero Mode Fidelity in the disordered Kitaev-Majorana chain}
\title{Strong zero modes in random Ising-Majorana chains}

\author{Saurav Kantha}
\email{skantha@irsamc.ups-tlse.fr}
\affiliation{Laboratoire de Physique Th\'{e}orique, Universit\'{e} de Toulouse, CNRS, France}
\author{Nicolas Laflorencie}
\email{nicolas.laflorencie@cnrs.fr}
\affiliation{Laboratoire de Physique Th\'{e}orique, Universit\'{e} de Toulouse, CNRS, France}

\date{\today}

\begin{abstract}
We investigate the fate and robustness of topological strong zero modes (SZMs) in random Ising–Majorana chains using the SZM fidelity, ${\cal F}_{\rm SZM}$, as a many-body diagnostic that quantifies how accurately SZM operators map the {\it entire} spectrum between opposite parity sectors. In clean systems, ${\cal F}_{\rm SZM}=1$ in the topological phase, vanishes in the trivial regime, and takes the universal value $\sqrt{8}/\pi$ at the $(1+1)$D Ising critical point. Here we study how quenched disorder modifies this picture across the infinite-randomness fixed point (IRFP) governing the criticality of the random chain. 
In both microcanonical and canonical ensembles, SZMs persist throughout the topological phase, including the gapless Griffiths regime, with fidelities converging exponentially to unity. At the IRFP, however, the fidelity distributions become ensemble dependent: the microcanonical ensemble displays bimodal peaks at $\{0.5,1\}$, while the canonical ensemble develops a triple-peak structure at $\{0,0.5,1\}$ with power-law singularities. 
Our results establish ${\cal F}_{\rm SZM}$ as a robust probe of localization-protected topological order and uncover distinctive topological features of infinite-randomness criticality. Unlike the clean Ising CFT, where the finite critical value arises from a cancellation of power laws, the IRFP seems to exhibit an intrinsically stronger topological character. The edge-selective structure of the critical distributions may suggest a boundary manifestation of the average Kramers–Wannier duality symmetry at the IRFP.
\end{abstract}

\maketitle

%\tableofcontents

\section{\label{sec:introduction}Introduction}
\subsection{Generalities}
Topological Majorana zero modes, protected by an underlying symmetry have been a subject of intense study over the past two decades, from theoretical as well as experimental directions~\cite{Lutchyn2010, Oreg2010, Sarma2015, MiSonnerNiuEtAl2022,Marra2022, Dvir2023,jin_topological_2025,rachel_majorana_2025}, primarily because of the non-Abelian nature of Majorana exchange, that offers a path towards fault-tolerant quantum computing~\cite{Kitaev2001,NonAbelianStatisticsTopological2011,IvanovNonAbelian,AasenMajoranaTopoComputation2016}.

Majoranas, which are anti-particles of their own, form a charge-neutral subset of fermions. They correspond to solutions of the Dirac equation that are equal to their charge conjugates, resulting in the absence of a conserved $U(1)$ charge. Two such (real) Majorana fermions can be paired to form a (complex) Dirac fermion. While fundamental Majorana particles have not been detected in nature, Majorana-like quasiparticles have emerged in various quantum condensed matter platforms~\cite{Alicea2012, Aguado2017,OregRefaelOppen2010,FuKane2008}, primarily in topological superconductors~\cite{Beenakker2013,SatoAndo2017,Kitaev2001, Mourik2012}. In such systems with open boundaries and a non-zero superconducting gap, Majorana zero modes appear as quasiparticle excitations at zero energy, and are exponentially localized at the boundaries of the system.  

One of the largely studied models that hosts Majorana zero modes is the celebrated Kitaev chain~\cite{Kitaev2001}, which is a one-dimensional spinless fermionic p-wave superconductor described by the following free-fermion Hamiltonian
\begin{equation}
    {\cal{H}}_K = -\sum_{j}\left(\left[t_{j}^{\dagga} c^{\dagger}_jc^{\dagga}_{j+1} + \Delta_{j}^{\dagga} c^{\dagger}_j c^{\dagger}_{j+1} + \text{h.c.}\right]-\mu_{j}^{\dagga} c^{\dagger}_j c^{\dagga}_{j}\right).
    \label{eq:Kitaev_hamiltonian}
\end{equation}
Here $t_j, \Delta_j$ and $\mu_j$ are nearest-neighbor hopping, superconducting pairing, and chemical potential respectively, which can all be site-dependent. The superconducting pairing term breaks particle number conservation, but the fermionic parity is a conserved quantity. The parity operator $\mathcal{P}=(-1)^{\sum_j n_j}$ ($n_j=c_{j}^{\dagger} c_{j}^{\dagga}$ is the density at site $j$) commutes with the Hamiltonian.
In the clean limit, where the parameters 
$\{t_j,\,\Delta_j,\,\mu_j\}$ are site-independent, for $|\mu|<2t$ and $|\Delta|\neq 0$, dubbed the topological phase, the spectrum of the Kitaev chain {\it{with open ends}} contains two near-zero eigenvalues which become exactly zero, and two-fold degenerate in the thermodynamic limit. It is impossible to get rid of these "zero modes" without closing the superconducting gap. The superconducting gap cannot be closed by any local, parity-preserving perturbation. Hence, Majorana zero modes in the Kitaev-Majorana chain are said to be protected by parity symmetry. As one increases $|\mu|$, a gap-closing topological phase transition occurs at $|\mu|=2t$ beyond which the system is in the trivial phase, where the zero energy states no longer exist. For a Kitaev chain with periodic boundary conditions, zero modes do not appear in the spectrum (except at gap-closing critical points). Instead, the phases are marked by a topological index called the winding number. For real $\Delta$, the Kitaev chain belongs to class BDI of the Altland-Zirnbauer classification of topological superconductors~\cite{AZclassification}. Class BDI in one dimension is characterized by a winding number that can take values in $\mathbb{Z}$. However, in its present form in Eq.~\eqref{eq:Kitaev_hamiltonian}, the values of the winding number are limited to $0$ and $1$, characterizing the trivial and topological phase respectively. Higher winding numbers can be obtained by adding longer range hopping terms~\cite{niu_majorana_2012,DeGottardiLongRangehoppingMajorana2013,AlecceExtendedKitaev2017}.

\subsection{The Ising-Majorana chain}
As said before, each (complex) Dirac fermion operator can be split into two (real) Majorana fermion operators $a_j$ and $b_j$, such that  
\be 
c_j = \cfrac{a_j + i b_j}{2}\quad {\rm and}\quad  c^{\dagger}_j = \cfrac{a_j-ib_j}{2}.
\ee 
 It is straightforward to verify that these new operators are their own antiparticle $(a/b)^{\dagger}=(a/b)$, square to the identity $(a/b)^2=1$, and satisfy the  anti-commutation rules $\{a_i,a_j\}=\{b_i,b_j\}=2\delta_{ij}$, and $\{a_i,b_j\}=0$.

Hence, the above Kitaev Hamiltonian Eq.~\eqref{eq:Kitaev_hamiltonian} can be rewritten in the Majorana language as follows 
\begin{equation}
    {\cal{H}}_M = i\sum_{j} \Big(X_j b_j a_{j+1} - Y_ja_{j}b_{j+1} + h_j a_j b_j\Big),
    \label{eq: Majorana_hamiltonian}
\end{equation}
with couplings $X_j=(t_j+\Delta_j)/2$, $Y_j=(t_j-\Delta_j)/2$, and $h_j=\mu_j/2$.
One can also introduce another language, using the standard Jordan-Wigner transformation 
\bea
a_j=K_j\sigma^x_j,\quad 
b_j=K_j\sigma^y_j,\quad
a_jb_j=i\sigma_j^z,
\eea
(with the non-local ``string'' $K_j=\prod_{k=1}^{j-1}\sigma_k^z$), to map the above Majorana chain model Eq.~\eqref{eq: Majorana_hamiltonian} to the paradigmatic transverse-field $XY$ spin chain Hamiltonian~\cite{LIEB1961407}
\begin{equation}
    {\cal H}_{XY} = -\sum_{j} \Big(X_j \sigma^x_j \sigma^x_{j+1}+ Y_j \sigma^y_j \sigma^y_{j+1} + h_j\sigma^z_j\Big),
    \label{eq: Hamiltonian_XY}
\end{equation}
which simply reduces in the limit $Y_j=0$ (which corresponds to $t_j=\Delta_j$) to the celebrated transverse-field Ising (TFI) chain~\cite{pfeutyOnedimensionalIsingModel1970}. Throughout the paper, we will work in this TFI limit. In the clean case, the ground-state of the TFI chain exhibits two phases, a disordered paramagnet at large field and a magnetic ordered phase at small field, separated by a quantum critical point at $h=X$, which exactly corresponds to the topological phase transition point $|\mu|=2t$ of the clean Kitaev chain. The topological and the trivial phases correspond to ordered and disordered phases, respectively, see Fig.~\ref{fig:sketch} (a). While an exact solution to the TFI chain was known for a long time, the topological properties of the corresponding fermionic Kitaev-Majorana chain were uncovered by Kitaev~\cite{Kitaev2001}, decades after the works of Lieb, Schultz, Mattis~\cite{LIEB1961407}, and Pfeuty~\cite{pfeutyOnedimensionalIsingModel1970}.   

In the rest of the work, we focus on the topological properties of the TFI chain in the general case of random couplings and fields, for finite chains of length $L$, dubbed from now random Ising-Majorana (RIM) which is equivalently written in terms of Pauli or Majorana operators as follows
\begin{equation}
\label{eq:RIM}
\begin{aligned}
\mathcal{H}_{\rm RIM}
&= -\sum_{j=1}^{L} \Big( X_j \sigma^x_j \sigma^x_{j+1}
    + h_j \sigma^z_j \Big) \\
&= i \sum_{j=1}^{L} \Big( X_j b_j a_{j+1}
    + h_j a_j b_j \Big) .
\end{aligned}
\end{equation}
with open boundary conditions (OBC), i.e. $X_L=0$. 

\subsection{Paper organization}

The remainder of this paper is organized as follows.
In Sec.~\ref{sec:SZMs} we first define the strong zero mode (SZM) operators that form the central object of our study. We review their key properties and discuss how disorder modifies their spatial structure.  We then introduce in Sec.~\ref{sec:F} the fidelity-based diagnostics that will be used throughout the paper to characterize the SZM robustness. After reviewing the disorder-free case for which exact analytical results are available for the SZM fidelity, we discuss the influence of randomness on the global phase diagram of Ising-Majorana chains, with a particular emphasis on Griffiths regimes and the infinite randomness criticality.

Sec.~\ref{sec:mc} contains our main numerical results for the microcanonical disorder ensemble. We first analyze the average and typical SZM fidelities in the trivial and topological phases, establishing their distinct finite-size scaling forms. At criticality, we characterize in detail the  entire distributions of fidelity and show that the IRFP exhibits a singular bimodal structure. In particular, the symmetrized fidelity develops a double-peak distribution at $\{1/2,1\}$, whose weights flow toward equal values, yielding an asymptotic critical fidelity ${\cal{F}}_{\mathrm{SZM}} \to 3/4$. We also identify the disorder-dependent crossover length scale governing the flow from the clean Ising fixed point to the IRFP.

In the final discussion part (Sec.~\ref{sec:disc}), we first address ensemble dependence by comparing microcanonical and canonical disorder sampling, and we then place our results in a broader context, including connections to recent works on average duality, as well as some possible experimental implications for Rydberg atoms. We finally conclude in Sec.~\ref{sec:conclusion} with a summary and an outlook.

\section{Theoretical framework}
 \subsection{Strong zero modes}
 \label{sec:SZMs}
Thanks to Fendley~\cite{fendleyParafermionicEdgeZero2012, fendleyStrongZeroModes2016}, and following works~\cite{kempLongCoherenceTimes2017,elsePrethermalStrongZero2017,olundBoundaryStrongZero2023,vernier_strong_2024,lariviere_exact_2025,gehrmann_exact_2025,essler_strong_2025}, it is now well known that for a certain class of integrable models (such as TFI or XYZ chains) with open boundaries, one can exactly construct operators which commute with the Hamiltonian up to small corrections, which exponentially vanish with the chain length. This property has a strong influence over the {\it{entire}} many-body spectrum, beyond low energy~\cite{Wada2021,alexandradinata_parafermionic_2016}. Such a Strong Zero Mode (SZM) operator $\Psi$ fulfils the following properties:

\begin{enumerate}
    \item\label{cond: szm_1} it commutes with the Hamiltonian up to an error that is exponentially small in the system size, i.e.  $\|[{\cal{H}},\Psi]\|<\exp(-L/\xi)$ for a finite positive $\xi$,
    \item\label{cond: szm_2} it anti-commutes with a discrete symmetry operator of the Hamiltonian (here the parity $\mathcal{P}$), and
    \item\label{cond: szm_3} it is normalizable in the thermodynamic limit, i.e $\Psi^2=1$.
\end{enumerate}

SZMs are more resilient than low-energy zero modes because, unlike the latter, they cannot be destroyed by thermal excitations from the bulk, making them robust to finite-temperature phenomena. Moreover, because they commute with the Hamiltonian up to an exponentially small correction in the system size, this gives the edge SZMs extremely long coherence times~\cite{fendleyStrongZeroModes2016, kempLongCoherenceTimes2017, elsePrethermalStrongZero2017}. In non-interacting systems like the TFI chain, the coherence time is infinite~\cite{Mila2018} in the thermodynamic limit. Fendley showed that this is true even for a certain class of interacting but integrable XYZ spin chain~\cite{fendleyStrongZeroModes2016}. In non-integrable systems however, one observes that the edge mode coherence persists for a finite, but extremely long duration~\cite{kempLongCoherenceTimes2017,Tausendpfund2025,Yates2020}. In such systems, the norm of the commutator of the SZM with the Hamiltonian is exponentially small in inverse of interaction strength, in addition to the usual finite size exponential correction.

SZM operators can be constructed for both, clean and disordered Ising-Majorana chains~\cite{huseLocalizationprotectedQuantumOrder2013}. For the RIM Hamiltonian of Eq.~\eqref{eq:RIM}, one can construct two such SZMs, $\Psi_{l}$ and $\Psi_{r}$ localized at the left and right boundaries of the chain respectively. They can be expressed as linear combinations of the Majorana fermion operators as follows
\begin{alignat}{2}
\Psi_{l} = \frac{1}{\mathcal{N}_l}\sum_{j=1}^L \Theta^a_{j-1}a_j &\qquad \Psi_{r} = \frac{1}{\mathcal{N}_r} \sum_{j=1}^L \Theta^b_{j-1}b_{L-j+1}
\label{eq:SZM_lr}
\end{alignat}
with  coefficients
\begin{alignat}{2}
  \Theta^a_{j>0} = \prod_{i=1}^j \frac{h_i}{X_i}       &\qquad  \Theta^b_{j>0} = \prod_{i=1}^j \frac{h_{L-i+1}}{X_{L-i}},
  \label{eq:thetas}
\end{alignat}
and boundary conditions $\Theta^a_0=\Theta^b_0=1$. 
The normalization factors
$\mathcal{N}_l$ and $\mathcal{N}_r$ are given by $\mathcal{N}_{l,r}^2 = \sum_{j=0}^{L-1} |\Theta^{a,b}_j|^2$, which can be expressed as sums of Kesten variables~\cite{monthusFinitesizeScalingProperties2004a}

\be
\label{eq:norms}
\begin{aligned}
\mathcal{N}_l^2 &= 1+\sum_{j=1}^{L-1}\prod_{i=1}^j \bigg(\frac{h_i}{X_i}\bigg)^2 \\
\mathcal{N}_r^2 &= 1+\sum_{j=1}^{L-1}\prod_{i=1}^j \bigg(\frac{h_{L-i+1}}{X_{L-i}}\bigg)^2.
\end{aligned}
\ee
The above normalization condition \ref{cond: szm_3} is satisfied for both the left and right SZM operators provided that the global control parameter of the RIM chain, $\delta$, is such that
\be
\delta={\overline{\ln X}}-{\overline{\ln h}}
\label{eq:delta}
\ee 
is positive, which corresponds to the ordered regime of the random TFI chain, as studied in pionner work~\cite{fisherRandomTransverseField1992,fisherCriticalBehaviorRandom1995,youngNumericalStudyRandom1996,igloiRandomTransverseIsing1998,fisherDistributionsGapsEndtoend1998}. Condition \ref{cond: szm_2} is obviously met, since each term in $\Psi_{l/r}$ anti-commutes with the parity operator 
\be
\mathcal{P}=\prod_{j=1}^{L}\sigma_j^z\quad \Rightarrow\quad \{\Psi_{l/r}, \mathcal{P}\} = 0.\ee
And finally, the commutators of $\Psi_{l/r}$ with the RIM Hamiltonian can straightforwardly be computed for each individual sample
\be
\label{eq: szm_hamiltonian_commutator}
\begin{aligned}
[{\cal{H}}_{\rm RIM},\Psi_{l}] &= -2i\cfrac{\Theta_{L-1}^{a}}{\mathcal{N}_l}\,b_L \\
[{\cal{H}}_{\rm RIM},\Psi_{r}] &= 2i\cfrac{\Theta_{L-1}^{b}}{\mathcal{N}_r}\,a_1.
\end{aligned}
\ee
We see that the norms typically vanish exponentially 
\be 
\|[{\cal{H}}_{\rm RIM},\Psi]\|\propto \cfrac{2X_{\rm typ}}{{\cal N}_{l,r}}\exp(-\delta L),
\ee 
where $X_{\rm typ}=\exp({\overline{\ln X}})$ is the typical coupling, and
the decay is controlled by the typical localization length of the SZMs in Eq.~\eqref{eq:thetas}: 
\be
\xi\equiv \frac{1}{\delta}.
\ee

\subsection{Fidelity: A many-body marker for SZMs}
\label{sec:F}
\subsubsection{Definitions}
The many-body spectrum of am Ising-Majorana chain can be split into two parity sectors due to the underlying $\mathbb{Z}_2$ symmetry. In the ordered regime $\delta>0$, the action of a SZM operator on the $n^{\rm th}$ eigenstate $\ket{n,p}$ with parity $p=\pm 1$, maps it to its (nearly degenerate) ``partner'' eigenstate $\ket{n,-p}$, which belongs to the opposite parity sector, see Ref.~\cite{fendleyParafermionicEdgeZero2012,laflorencieUniversalSignaturesMajorana2023}. In order to quantify how good this mapping is, one can define the SZM fidelities for left and right SZM operators as follows
%\begin{align}
%    \Psi_{l}\ket{n,p} = %\mathcal{F}_{l}\ket{n,-p} + \sum_{m\neq n}\alpha_m\ket{m,-p}\label{eq:left_szm_|n,p>} \\
%    \sum_m\alpha_m\ket{m,-p}\label{eq:left_szm_|n,p>} \\
%    i\mathcal{S}_n\Psi_{r}\ket{n,p} = \mathcal{F}_{r}\ket{n,-p} + \sum_{m\neq n}\gamma_m\ket{m,-p} \label{eq:right_szm_|n,p>}
%i\Psi_{r}\ket{n,p} = \sum_{m}\beta_m\ket{m,-p} \label{eq:right_szm_|n,p>},
%\end{align}
%with the 
%Here, the factor $i\mathcal{S}_n = i\text{ sgn}(E_{n,p}-E_{n,-p})$ is multiplied to $\Psi_r$ to make $\mathcal{F}_r$ real and positive. The normalization of $\Psi_{l,r}$ ensures $\mathcal{F}_{l,r}\in[0,1]$. The two quantities 
\be
   \mathcal{F}_{l,r} = \left|\bra{n,-p}\Psi_{l,r}\ket{n,p}\right|
   \label{eq:defFid}
\ee
Building on the commutation relations Eq.~\eqref{eq: szm_hamiltonian_commutator}, one can show (see App.~\ref{sec:app_szm}) that left and right fidelities are independent of the energy levels $n$, and can be simply expressed as functions of the parity gap (see Ref.~\cite{laflorencieUniversalSignaturesMajorana2023}) 
\be
\Delta_p=\left|\bra{n,-p}{\cal{H}}_{\rm RIM}\ket{n,-p}-\bra{n,p}{\cal{H}}_{\rm RIM}\ket{n,p}\right|,\ee 
the surface magnetizations 
\be 
m^s_{l,r} = |\bra{n,-p}\sigma^x_{1,L}\ket{n,p}|,
\label{eq:ms}
\ee
and the norms ${\cal{N}}_{l,r}$ from Eq.~\eqref{eq:norms}, such that for each sample, one can write
\be
    \mathcal{F}_{l,r} = 2\cfrac{m_{r,l}^{s}}{\mathcal{N}_{l,r}}\,\cfrac{\Theta_{L-1}^{a,b}}
    {\Delta_p}.
    \label{eq:Flrdis}
\ee
These two fidelities can therefore serve as many-body markers for the presence or absence of SZMs in the chain. One can also symmetrize them \cite{laflorencieUniversalSignaturesMajorana2023} to form 
\bea
        \mathcal{F}_{\rm SZM} &=& \frac{1}{2}(\mathcal{F}_l +\mathcal{F}_r)\nonumber\\
        &=& \frac{1}{\Delta_p}\left(\cfrac{m_{l}^{s}\Theta_{L-1}^{b}}{\mathcal{N}_{r}}+\cfrac{m_{r}^{s}\Theta_{L-1}^{a}}{\mathcal{N}_{l}}\right).
        \label{eq: symmetrized_fidelity_defn}
\eea
Since $\mathcal{F}_{l,r}\in[0,1]$, $\mathcal{F}_{\rm SZM}$ also lies in the same interval. The closer this number is to $1$, the better the mapping between the two parity sectors. 

\begin{figure}[b!]
    \centering
    \includegraphics[width=\linewidth]{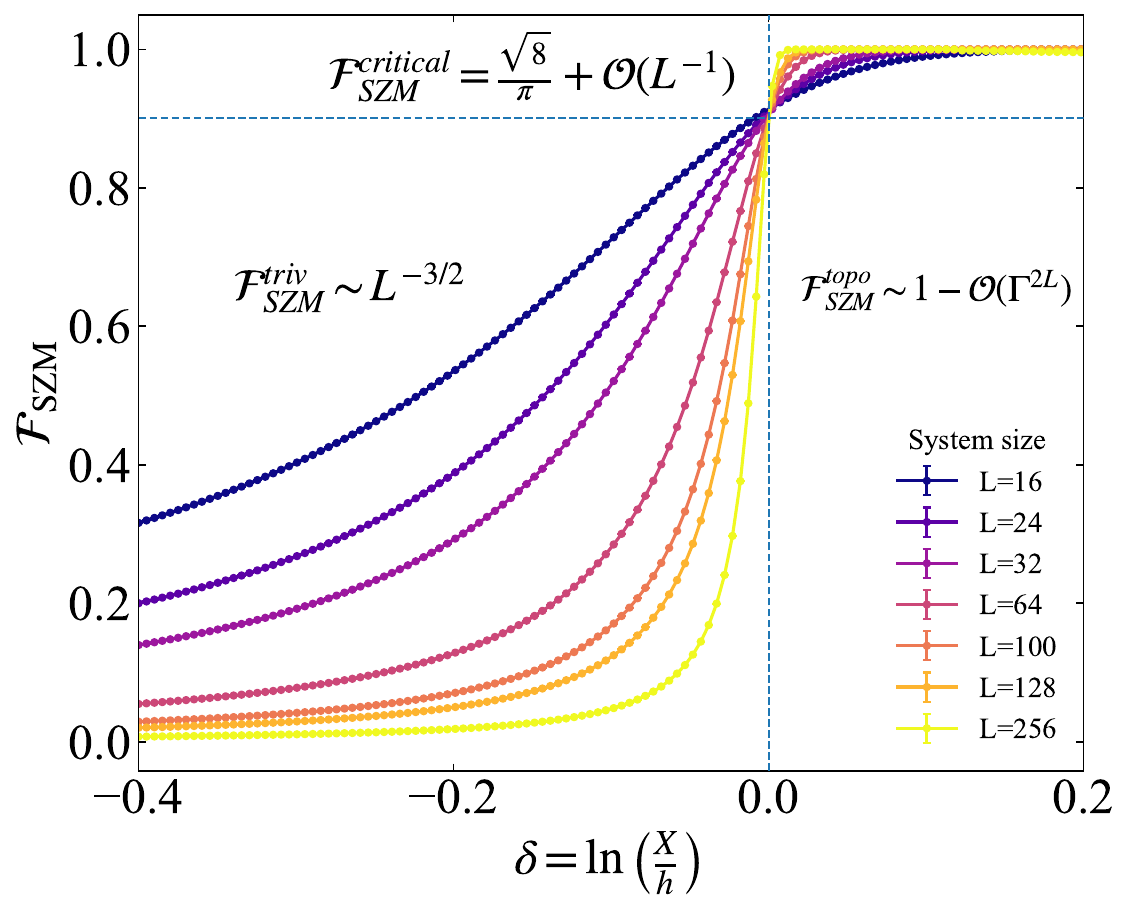}
    \caption{\label{fig:clean}Symmetrized fidelity for the clean Ising-Majorana (Kitaev) chain. Different curves correspond to different system sizes indicated in the legend. The vertical dotted line indicates the critical point and the horizontal dotted line indicates the universal value $\sqrt{8}/\pi$ that $\mathcal{F}_{\rm SZM}$ takes at the Ising criticality.}
\end{figure}

\subsubsection{Disorder-free case}
The usage of fidelity as a marker of SZMs is easily observed in the clean case where it goes from a value $1$ to $0$ from a topological to trivial phase with an eigenstate phase transition at the $(1+1)D$ Ising critical point, where remarkably, its value becomes universal~\cite{laflorencieUniversalSignaturesMajorana2023}. At the critical point, SZM fidelity indicates an eigenstate phase transition, which affects the entire spectrum rather than just the ground state.

Before discussing in great detail the random model, we first summarize the results for the disorder-free case obtained in \cite{laflorencieUniversalSignaturesMajorana2023}.
For the clean chain, $h_j = h>0$ and $X_j = X$, $\forall j$. The left and right SZM operators can be expressed as follows, using $\Gamma = {h}/{X}$,
\begin{equation}
            \Psi_{l} = \sum_{j=1}^L \Gamma^{j-1} a_j\quad \quad 
        \Psi_r = \sum_{j=1}^L \Gamma^{j-1}b_{L-j}.
\end{equation}
Hence, the commutators with the Hamiltonian are
\begin{alignat}{2}
    [H,\Psi_{l}] = -\frac{2iX\Gamma^L}{\mathcal{N}_0}b_L  &\qquad [H,\Psi_{r}] = \frac{2iX\Gamma^L}{\mathcal{N}_0}a_1, \label{eq: clean_szm_hamiltonian_commutator}
\end{alignat}
with normalization constant $\mathcal{N}_0 =\sqrt{(1-\Gamma^{2L})/(1-\Gamma^2)}$. $\Gamma<1~(>1)$ corresponds to the topological (trivial) phase respectively, both of which are gapped phases for the low-energy bulk excitations. In the topological phase, the SZM fidelity is equal to $1$ with exponentially small finite size corrections: 
\be
\mathcal{F}^{\rm topo}_{\rm SZM} = 1-\mathcal{O}(\Gamma^{2L}),
\label{eq:Fcleantopo}
\ee
whereas in the trivial phase it vanishes as a power-law
\be 
\mathcal{F}^{\rm trivial}_{\rm SZM}\sim L^{-3/2}.
\label{eq:Fcleantrivial}
\ee
At the Ising critical point $\Gamma=1$ in the clean chain, where the correlation length $\xi_{\rm clean} = \left|\ln\Gamma\right|^{-1}$ diverges, the zero modes no longer exist; instead they take the form of gapless plane-wave solutions penetrating the entire bulk. 
Despite this, surprisingly the SZM fidelity does not vanish at this point; rather it takes a universal value 
\be
\mathcal{F}^{\rm ISING}_{\rm SZM} = \sqrt{8}/\pi\approx 0.90032,
\label{eq:FcIsing}
\ee 
with leading order finite size corrections $\sim L^{-1}$. Moreover, the universality of the SZM fidelity also manifests for the transverse-field $XY$ model (Eq.~\eqref{eq: Hamiltonian_XY} corresponding to a Kitaev chain away from $t=\Delta$). Along the entire critical line $h = X+Y$, the critical fidelity takes the universal value of Eq.~\eqref{eq:FcIsing}.

%%%%%%%%%%%%%%%%%%%%%%%
\begin{figure*}
       \centering
\includegraphics[width=1.5\columnwidth,clip]{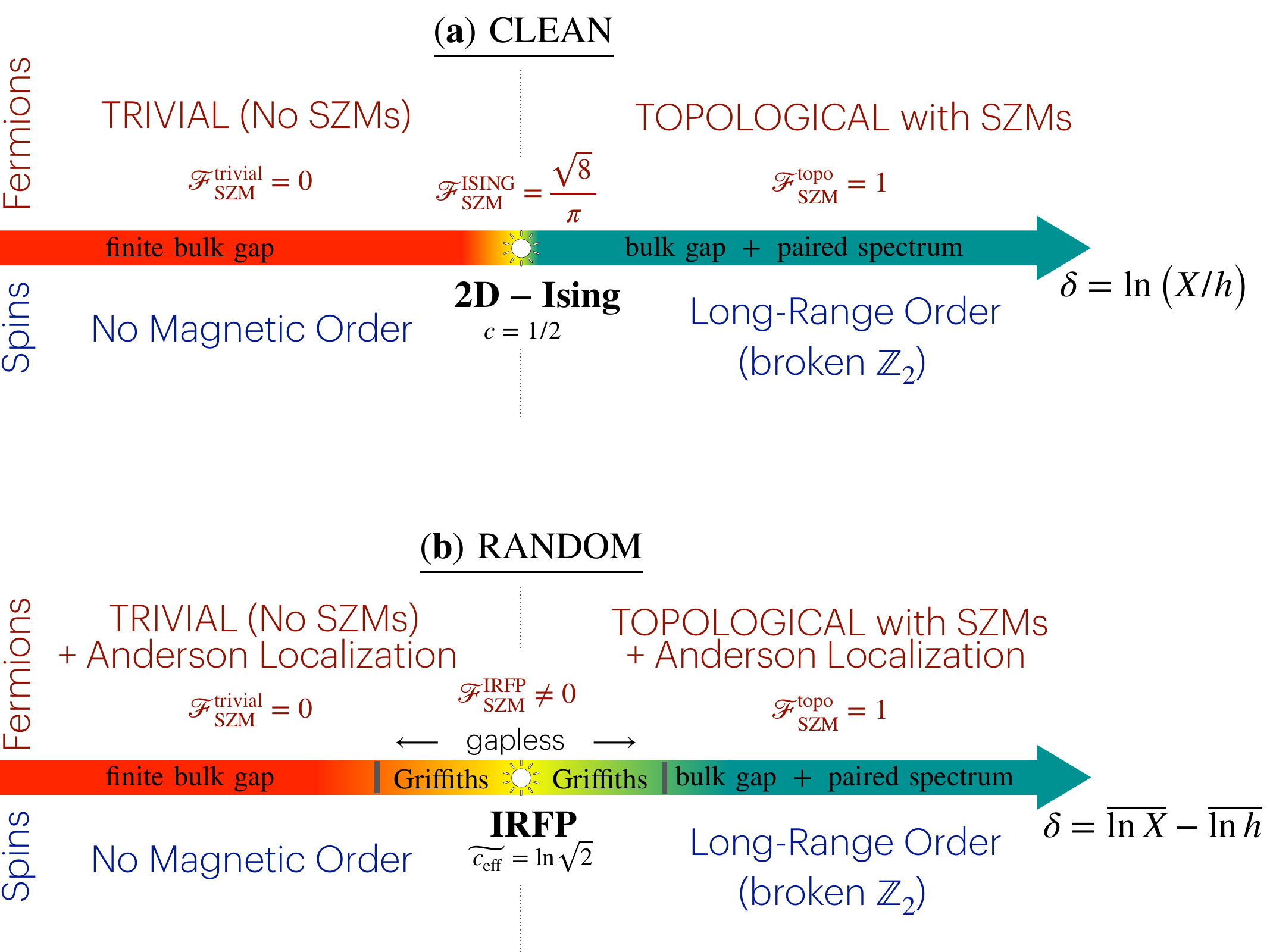}
           \caption{\label{fig:sketch}\textbf{Overview of the phase diagram of clean and random Ising--Majorana chains.}
(a) {\it{Clean system:}} The (1+1)D Ising critical point (central charge $c=1/2$) separates a topological phase supporting SZMs from a trivial phase without SZMs. In the topological phase, a finite bulk gap coexists with parity-paired many-body spectra and unit fidelity, at large enough sizes $\mathcal{F}_{\rm SZM}\to 1$, while in the trivial phase $\mathcal{F}_{\rm SZM}\to 0$. At criticality, the fidelity takes the universal value $\mathcal{F}_{\rm SZM}^{c}=\sqrt{8}/\pi$~\cite{laflorencieUniversalSignaturesMajorana2023}. In the spin representation, the ordered phase exhibits broken $\mathbb{Z}_2$ symmetry.
(b) {\it{Random case:}} Disorder drives the Ising transition to an infinite-randomness fixed point (IRFP) characterized by an {\it{effective central charge}}
${\widetilde{c_{\rm eff}}}=\ln \sqrt{2}$. The topological phase persists in the presence of Anderson localization and extends into a gapless Griffiths regime. Throughout this phase, SZMs remain robust with $\mathcal{F}_{\rm SZM}\to 1$ in the thermodynamic limit, while the trivial phase remains non-topological with zero SZM fidelity. At criticality, the IRFP displays non-trivial fidelity distributions, depending on the disorder ensemble, as discussed in the text, yielding a finite fidelity at the IRFP.}
\end{figure*}

%%%%%%
\subsection{Influence of the disorder on the phase diagram of the Ising-Majorana chain}
\label{sec:dis_PD}

In the presence of disorder, either in the couplings $X_j$ or in the fields $h_j$, the zero-temperature phase diagram and the nature of the quantum phase transition in the TFI chain are significantly altered, as established in the pioneering works~\cite{mccoy_theory_1968,fisher_random_1992,fisherCriticalBehaviorRandom1995}. In particular, most low-energy properties have been described by Fisher~\cite{fisher_random_1992,fisherCriticalBehaviorRandom1995} using the strong disorder renormalization group (SDRG) method (see Ref.~\cite{igloiStrongDisorderRG2018} for a recent review). Fig.~\ref{fig:sketch} gives an overview of the ground-state phase diagram, comparing clean and random cases, with control parameter $\delta$, see Eq.~\eqref{eq:delta}. We briefly recall the main properties below.

\subsubsection{Disordered-trivial regime}
For $\delta<0$, clean and random cases present similar properties sufficiently deep in the phase, i.e. a disordered paramagnetic ground-state, with a finite gap far enough from the critical point. In terms of fermions, the disordered model displays Anderson localization. When the critical point is approached, there is an additional gapless Griffiths regime (dominated by rare regions~\cite{griffiths_nonanalytic_1969}) where the finite-size scaling of the lowest energy gap is algebraic with the chain length $L$~\cite{fisher_phase_1999,motrunich_griffiths_2001,igloiExactRenormalizationRandom2002}, $\Delta\sim L^{-z}$, controlled by a non-universal dynamical exponent $z\approx |2\delta|^{-1}$.

\subsubsection{Ordered-topological regime} 
The other side of the phase diagram ($\delta>0$) hosts the ordered regime, where spins exhibit long-range order in the bulk, $\langle \sigma_i^x \sigma_{i+r}^x \rangle \neq 0$, when  $r\to\infty$. In the thermodynamic limit, the $\mathbb{Z}_2$ symmetry is spontaneously broken, with a twofold-degenerate ground state, one in each parity sector. As in the disordered-trivial  phase, a Griffiths regime also appears on the ordered side in the vicinity of the critical point, characterized by gapless excitations that are Anderson-localized in the bulk. 
Importantly, the topological properties of the ordered phase persist in the presence of randomness: open chains support strong zero modes (SZMs) localized at the edges, whose existence and robustness extend throughout the ordered phase, including the Griffiths regime. For finite system size $L$, the two nearly degenerate ground states are separated by the exponentially small parity gap
$\Delta_p \sim \exp(-L/\xi)$,
with a characteristic length scale $\xi = 1/|\delta|$. It is important to note that $\Delta_p$ vanishes parametrically faster with increasing $L$ than the bulk gap associated with bulk Griffiths rare-region excitations.

\subsubsection{Quantum criticality: infinite randomness} For $\delta = 0$, the clean $(1+1)$D Ising critical point, which is described by a conformal field theory with central charge $c = 1/2$, is radically altered by the presence of any finite randomness. This was first demonstrated using the asymptotically exact strong-disorder renormalization group (SDRG)~\cite{fisherRandomTransverseField1992,fisherCriticalBehaviorRandom1995} and subsequently confirmed by numerical studies~\cite{youngNumericalStudyRandom1996,igloiRandomTransverseIsing1998,fisherDistributionsGapsEndtoend1998}. The quantum critical behavior obeys  the Infinite Randomness Fixed Point (IRFP) physics, which can be seen as a (non-interacting) quantum glass that displays marginal localization in terms of fermions~\cite{nandkishore_marginal_2014}. The IRFP exhibits several remarkable attributes, like absence of self-averaging for some key observables. The low energy gap $\Delta$ has a broad distribution even on a logarithmic scale; hence its average and typical values scale differently: for a chain of length $L$, the typical gap $\overline{\ln \Delta} \propto -L^{\psi}$, with $\psi=1/2$. This results in activated scaling characterized by the dynamical critical exponent $z=\infty$. Spin correlation functions also lack self-averaging: while the average spin-spin correlation has a power-law decay, typical correlations decay much faster as stretched exponential~\cite{fisherCriticalBehaviorRandom1995}.

More recently, there has been increased interest in the entanglement properties at the IRFP, motivated by the unexpected scaling of the disorder-averaged entropy with the block length~$\ell$ ($S^{\rm IRFP}_{\ell}={\widetilde{c_{\rm eff}}}\,\ln \ell + \text{constant}$),
where ${\widetilde{c_{\rm eff}}}=\ln\sqrt{2}=c\ln 2$ morally lays the role of an {\it{effective central charge}}~\cite{Refael2004PRL,Laflorencie2005PRB,Santachiara2006JSTAT,Hoyos2007PRB,Igloi2008JSTAT,Refael2009JPA,Fagotti2011PRB}, in echoes to the clean case~\cite{holzhey_geometric_1994}. However, the analogy is limited, as the underlying physical origin is markedly different: the logarithmic growth arises from the statistical counting of rare, long-range, entangled spin pairs which strongly contribute to the entropy, the rest of the system being almost not entangled. It is worth noting that this IRFP feature remains true beyond ground-state physics, at {\it{all anergies}}~\cite{Laflorencie_2022}.

\subsection{Topological properties: questions and main results}

The problem of Majorana chains in the presence of randomness is of central importance and has attracted sustained interest over the years~\cite{brouwer_probability_2011,bravyi_disorder_2012,huseLocalizationprotectedQuantumOrder2013,altlandTopologyAndersonLocalization2015a,DasSarmaPan2021_PRB_DisorderZBP,DeGottardiLongRangehoppingMajorana2013,hegdeMajoranaWavefunctionOscillations2016,Gergs2016,chepigaResilientInfiniteRandomness2024}. Nevertheless, the specific interplay between SZMs and disorder has received relatively limited attention~\cite{muller_classical_2016,Monthus2018,Lieu2018,Francica2023,peeters_effect_2024}. In their seminal work~\cite{huseLocalizationprotectedQuantumOrder2013}, Huse and co-workers discussed the robustness of quantum (topological) order against disorder, even in the absence of a bulk gap, for instance in the Griffiths regime approaching an IRFP. A key aspect of this problem is that Anderson localization of the fermionic degrees of freedom persists at all energies, thereby protecting the quantum order throughout the whole many-body spectrum, at least in the absence of interaction. The fate of such {\it{infinite-temperature}} localization-protected topological order in the presence of finite interactions—e.g., putative many-body localized (MBL) systems with topological order—has also been numerically investigated in a few works~\cite{Hannukainen_2024,artiaco_universal_2024,Kells2018PRB,Sahay2021PRL,Wahl2022PRB,Kjall2014PRL,Carmele2015PRB,Bahri2015NatComm,Moudgalya2020arXiv,laflorencieTopologicalOrderRandom2022}. Despite numerical evidence of spectral pairing in MBL phases~\cite{laflorencieTopologicalOrderRandom2022}, whether MBL can protect SZM operators remains an open question.

Here, we adopt a slightly different perspective  
and investigate, in non-interacting systems, the robustness of SZM operators against weak and strong randomness. We focus on how this robustness can be quantitatively characterized in the presence of disorder and analyze the influence of infinite-randomness criticality on SZMs.
To this end, we target the disordered Ising–Majorana chain Hamiltonian Eq.~\eqref{eq:RIM}, and explore the properties of SZMs both in the vicinity of, and exactly at, the IRFP critical point. Building on the concept of Majorana zero-mode fidelities~\cite{laflorencieUniversalSignaturesMajorana2023}, introduced above in Sec.~\ref{sec:F}, see Eq.~\eqref{eq:defFid}, we employ this quantity as a topological marker that quantifies how faithfully the operators $\Psi_l$ and $\Psi_r$ are to map even-parity to odd-parity states for the {\it{entire}} many-body spectrum.
More specifically, we aim to address the following questions:

    (i) Away from criticality, in the  vicinity of the transition, how does the gapless Griffiths regime affect SZMs and the scaling behavior of the fidelities\,?

    (ii) What is the critical behavior of ${\cal F}_{\rm SZM}$ at the IRFP\,? How does it depend on the disorder sampling, and what are the resulting distributions\,?

    (iii) How does the system crossover from the clean Ising to the IRFP\,? What length scale controls this flow\,?

    (iv) Can our results be interpreted in terms of an average duality symmetry at the IRFP\,?

\begin{figure*}
  \centering
   \includegraphics[width=.9\textwidth, trim=0cm 0 0cm 0, clip]
    {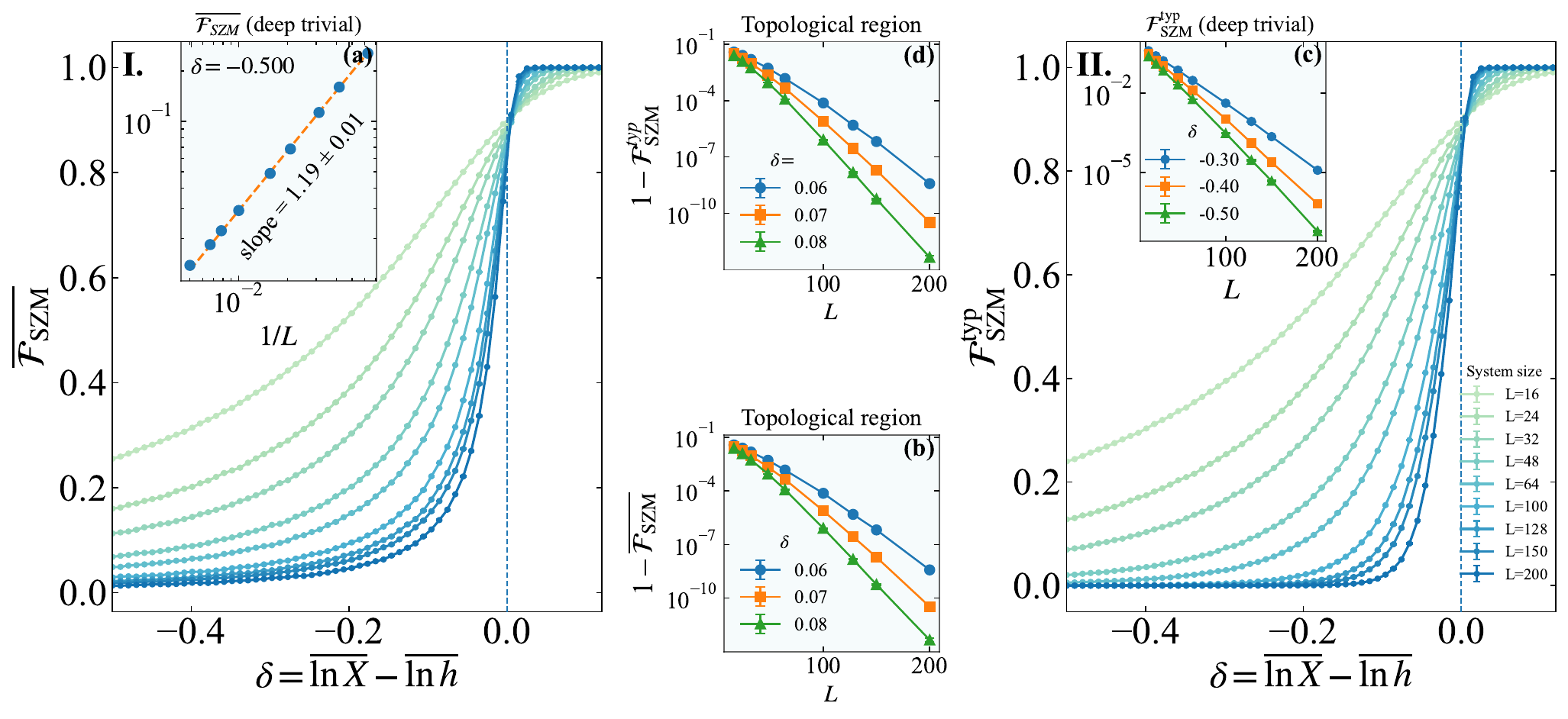}
  \caption{\label{fig:fidelity_full_phase_with_finite_size_corrections_mCan}
  Microcanonical ensemble results for the symmetrized fidelity for various system sizes, averaged over $2\times 10^4$ samples. {\bf{(I)}} Average ${\overline{{\cal{F}}_{\rm SZM}}}$ plotted against the control parameter $\delta$. The vertical dotted line marks the quantum critical point. Trivial and topological regions correspond to $\delta<0$ and $\delta>0$ respectively. Inset (a): log-log plot of $\overline{\mathcal{F}_{\rm SZM}}$ vs inverse system size, plotted for $\delta=-0.5$ (deep in the trivial region) showing a power-law decay to zero Eq.~\eqref{eq:Ftrivialpower-law}. Inset (b): $1-\overline{\mathcal{F}_{\rm SZM}}$ plotted against $L$ on a semi-log plot for various points in the topological region, showing exponential corrections to $\overline{\mathcal{F}_{\rm SZM}}$ as it approaches $1$.
  {\bf{(II)}} Same as plot I, for typical values ${\cal{F}}_{\rm SZM}^{\rm typ}$. Inset (c): Semi-log plot of $\mathcal{F}^{\rm typ}_{\rm SZM}$ for various values of $\delta$ deep in the trivial region, showing a exponential decay to zero. Inset (d): Semi-log plot of $1-\mathcal{F}^{\rm typ}_{\rm SZM}$ at inside the topological region, showing exponential finite size corrections.}
  \label{fig:wide}
\end{figure*}

\section{Results for the microcanonical Ensemble}
\label{sec:mc}
It is known~\cite{chayes_finite-size_1986,pazmandi_revisiting_1997} that the scaling properties of some observables may depend on the choice of ensemble for the random variables, as shown for the random Ising Majorana chain~\cite{igloiRandomTransverseIsing1998,dharEnsembleDependenceRandom2003,monthusFinitesizeScalingProperties2004a}.
In this work, we consider both canonical and microcanonical ensembles. However, here we focus on the microcanonical results in detail, while the canonical results are presented in App.~\ref{app:can}.

The microcanonical ensemble is distinguished from the canonical by fixing the value of $\delta = \overline{\ln X}-\overline{\ln h}$ at the individual level of {\it{each sample}} in the ensemble.
Observables such as the average and typical values of the lowest energy gap, surface magnetization, and end-to-end spin correlations have a strong dependence on the choice of ensemble. This is because the fluctuations of $\delta$, which slowly decay as $1/\sqrt{L}$ in the canonical ensemble for a chain of length $L$, are by definition absent for the microcanonical disorder ensemble.

In this section we present numerical results and discuss the behaviour of fidelities for the disordered Ising-Majorana chain with the microcanonical ensemble constraint. We perform exact diagonalization of the Hamiltonian Eq.~\eqref{eq:Kitaev_hamiltonian} written in particle-hole, or Nambu basis (see details in App.~\ref{sec:app_diago}).
The parameters couplings $\{X_j\}$ and fields $\{h_j\}$ are uniformly distributed such that
\begin{equation}
    \begin{split}
        & \overline{\ln X}=1\quad {\rm and}\quad 
         \overline{\ln h}=1-\delta
    \end{split}
\end{equation}
is satisfied for each sample and both $\{X_j\}$ and $\{h_j\}$ are in the intervals $[\overline{X}-W/2,\overline{X}+W/2]$ and  $[\overline{h}-W/2,\overline{h}+W/2]$ respectively, with width $W\geq 0$ controlling the disorder strength.

\subsection{Fidelity in the Topological and Trivial Regions}
We begin by presenting the properties of the symmetrized fidelity $\mathcal{F}_{\rm SZM}$ Eq.~\eqref{eq: symmetrized_fidelity_defn} in both trivial and topological regions. By trivial and topological regions, we refer to regions corresponding to $\delta<0$ and $\delta>0$ respectively. We keep the discussion of fidelity at the critical point $\delta=0$ for Sec.~\ref{sec:criticalF}. Nevertheless we present in Fig.~\ref{fig:fidelity_full_phase_with_finite_size_corrections_mCan} an overview of both average and typical values across the entire RIM phase diagram.

\subsubsection{Trivial Regime} 
Away from the critical point, in the trivial regime ($\delta<0$), both the average and the typical values of the SZM fidelity approach zero, as expected, but with different finite-size scaling forms. The average exhibits a power-law decay
\be 
{\overline{{\cal{F}}_{\rm SZM}^{\rm trivial}(L)}}\sim L^{-\omega},\quad \omega\approx 1.2,
\label{eq:Ftrivialpower-law}
\ee
similar to the disorder-free case~\cite{laflorencieUniversalSignaturesMajorana2023}, see Fig.~\ref{fig:clean} and Eq.~\eqref{eq:Fcleantrivial}. Note that we numerically observe that $\omega$ shows a weak dependence on $\delta$ far away from criticality, but close to the critical point, it varies rapidly, showing an apparent departure from power-law behaviour.

\begin{figure}[h!]
    \centering
    \includegraphics[width=1\linewidth]{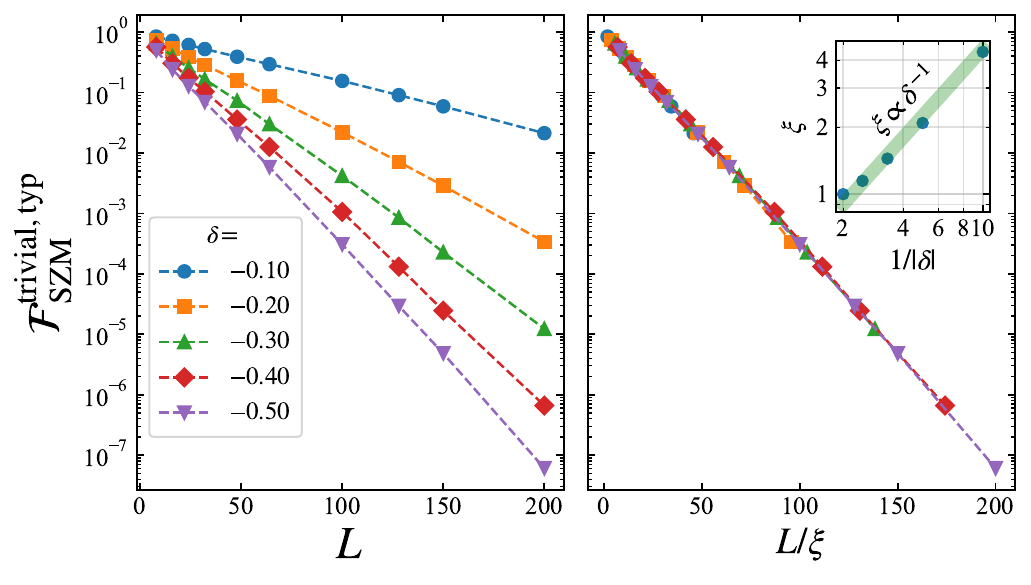}
    \caption{\label{fig:scalingFvsLoverxi_mc} (a) Exponential decay of the typical fidelity, plotted against $L$ for various values of $\delta<0$ in the trivial regime. (b) Data collapse when rescaling the length $L\to L/\xi$, where $\xi\sim |\delta|^{-1}$, see inset.}
\end{figure}

On the other hand, the typical fidelity exhibits a much faster decay to zero
\be 
{\cal{F}}_{\rm SZM}^{\rm trivial,\,typ}(L)\sim \exp(-L/\xi)
\label{eq:Ftrivialexp}
\ee
as visible in Fig.~\ref{fig:fidelity_full_phase_with_finite_size_corrections_mCan} (c) and Fig.~\ref{fig:scalingFvsLoverxi_mc}, with a length scale $\xi\propto |\delta|^{-1}$, which is the typical correlation length behavior~\cite{fisher_random_1992}.

\subsubsection{Topological Regime} 

\begin{figure*}
    \centering
    \includegraphics[width=.7\linewidth]{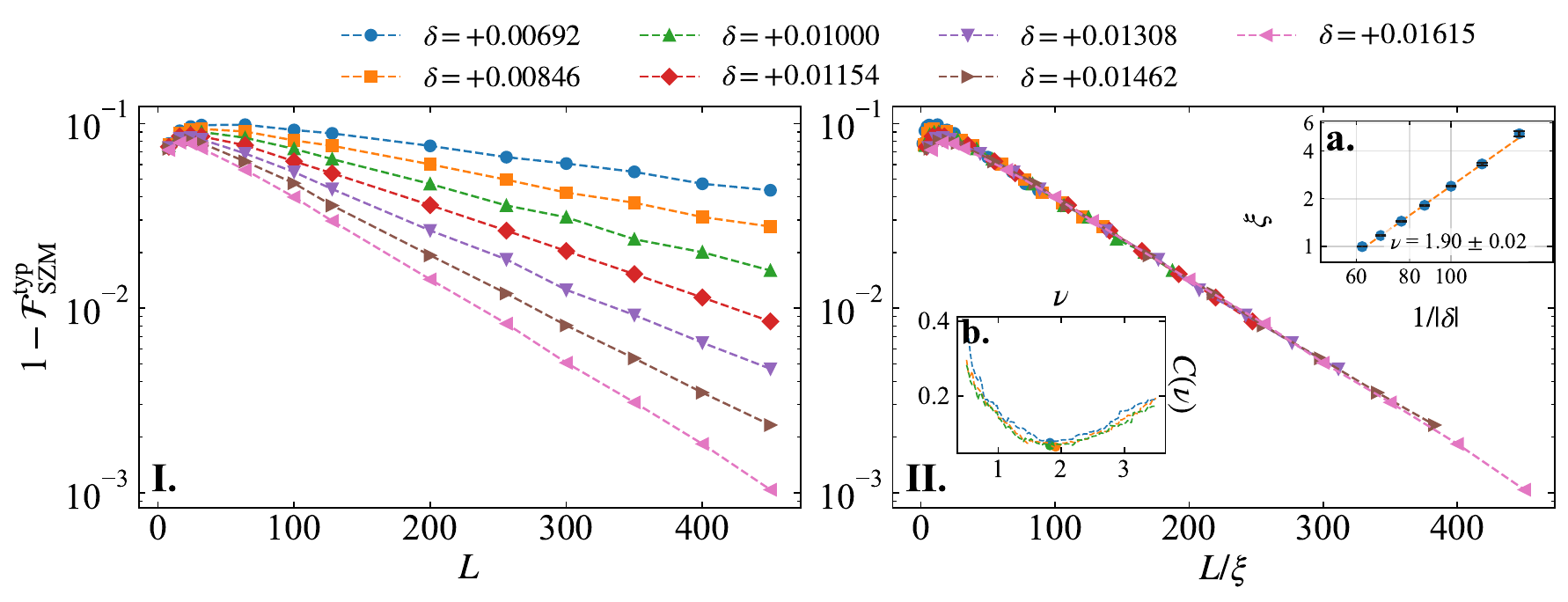}\hfill
    \includegraphics[width=.82\linewidth]{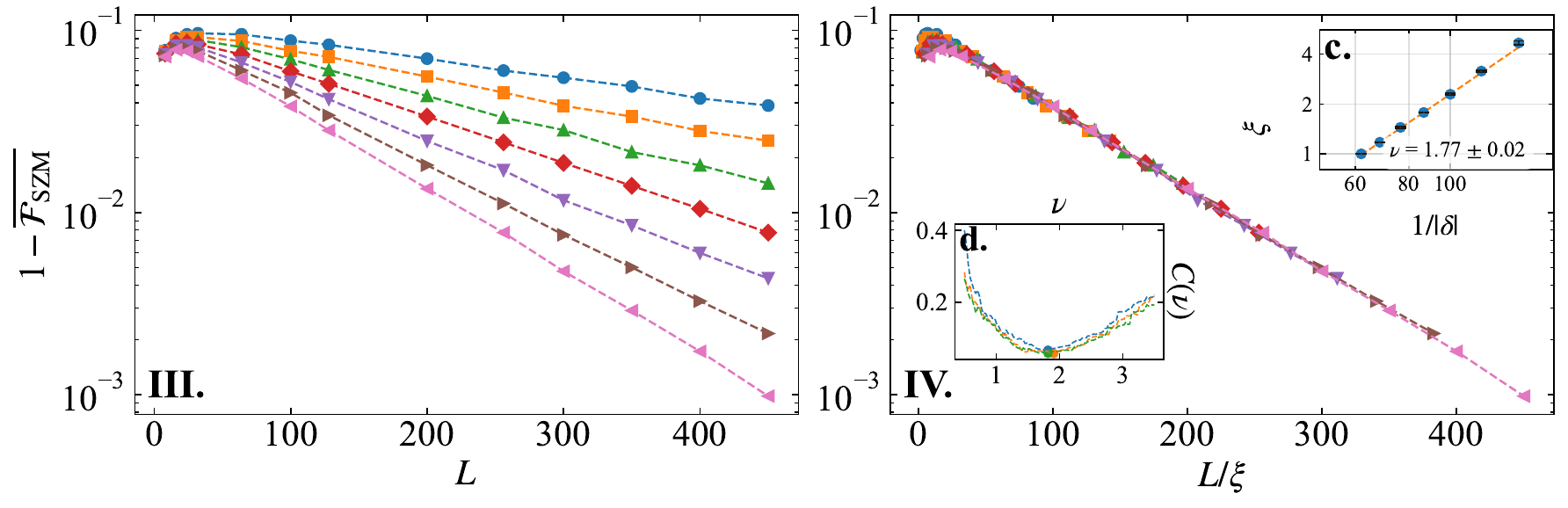}
    \caption{\label{fig:critical_exponent_topological_mCan}{\bf{(I)}}  Convergence of $\mathcal{F}_{\rm SZM}^{\rm typ}$ to $1$ in the topological regime, close to criticality, for various values of $\delta$ on a semi-log plot. {\bf{(II)}}  Collapse of the curves shown in {\bf{(I)}}  using one parameter scaling $L\rightarrow L/\xi$.  Inset (a): Length scale $\xi$ vs $1/|\delta|$ on a log-log plot showing a dependence $\xi \sim |\delta|^{-\nu}$. Inset (b): Values of the cost function $C(\nu)$ plotted against the critical exponent $\nu$. The different colors correspond to number of bins used to calculate $C(\nu)$ and the solid circles mark the minimum of the $C(\nu)$ curve. Both the insets show the value of $\nu\sim 2$.
    Panels {\bf{(III-IV)}} and insets (c) and (d) are the corresponding quantities for the average $\overline{\mathcal{F}_{\rm SZM}}$, which also exhibits a critical exponent $\nu$ close to $2$.}
    \end{figure*}
    
In the topological regime ($\delta>0$), both the average and typical values of ${\cal{F}}_{\rm SZM}$ approach $1$ in the same way, exponentially fast in system size, see Figs.~\ref{fig:fidelity_full_phase_with_finite_size_corrections_mCan} (b,d) and Fig.~\ref{fig:critical_exponent_topological_mCan}
\be
1-{\cal{F}}_{\rm SZM}^{\rm topo}\sim \exp(-L/\xi).
\ee
This is reminiscent of the clean case Eq.~\eqref{eq:Fcleantopo} where the associated length scale was $\xi_{\rm clean}= 1/\delta$, while in the disordered case we get
\be 
\xi(\delta)\sim \delta^{-\nu},
\ee 
with the same critical exponent for both typical and average, $\nu\approx 2$ that (surprisingly) seems to match the average exponent~\cite{fisher_random_1992}. 

One can extract the critical exponent in both cases by a single parameter scaling analysis of the curves $1-\mathcal{F}^{\rm typ}_{\rm SZM}$ and $1-\overline{\mathcal{F}_{\rm SZM}}$ at various values of $\delta$ close to $0^+$. 
Errors in the values of the scaling $\xi$ and the critical exponent $\nu$ in each case have been obtained by bootstrap with $1000$ samples. Fig.~\ref{fig:critical_exponent_topological_mCan} shows the result of our analysis. To verify the critical exponent that we obtain, we also calculate a cost function $C(\nu)$ that measures the mean squared deviation of the scaled points corresponding to the curves of different values of $\delta$. The precise definition and calculation of $C(\nu)$ is given in Appendix \ref{appendix:critical_exponent_details}. In both cases, typical and average, we find that the minimum of the curve $C(\nu)$ occurs for a value of $\nu$ that is very close to the one we obtain from single parameter scaling analysis, i.e. $\nu\approx 1.8-2$.

The above analysis shows that the SZM fidelities are asymptotically equal to one for any $\delta>0$, meaning that the SZM operators are robust and 
faithfully map even- to odd-parity many-body spectra. Importantly, this remains true even in the absence of a finite bulk gap, as it is the case in the topological Griffiths regime surrounding the IRFP. This being true for both sides of an open chain, at large size ${\cal{F}}_{l,r}\to 1$, we immediately recover the result of Igl\'oi and Rieger~\cite{igloiRandomTransverseIsing1998} from the definition Eq.~\eqref{eq:ms} that both (left and right) surface magnetizations $m^s_{l,r}$ are given by the inverse norms 

\be
\begin{aligned}
m^s_l&\to \bigg[1+\sum_{j=1}^{L-1}\prod_{i=1}^j \left(\frac{h_i}{X_i}\right)^2\bigg]^{-1/2} \\
m^s_r&\to \bigg[1+\sum_{j=1}^{L-1}\prod_{i=1}^j \left(\frac{h_{L+1-i}}{X_{L-i}}\right)^2\bigg]^{-1/2},
\end{aligned}
\ee
when $L$ is large. We can also get a closed expression for the finite-size scaling of the parity gap, using Eq.~\eqref{eq:Flrdis}
\be
\Delta(L)\approx 2m_l^s m_r^s \cfrac{\prod_{i=1}^L{h_i}}{\prod_{i=1}^{L-1}{X_i}},
\ee
which for a microcanonical disorder distribution exactly yields for $\delta>0$
\be
\Delta(L)\approx 2X_{\rm typ} m_l^s m_r^s\exp(-\delta L),
\ee
at the level of each individual sample. In this way, by relying solely on the SZMs, we recover the Fisher-Young relation that connects the bulk gap with boundary spins~\cite{fisherDistributionsGapsEndtoend1998}.

\subsection{Critical fidelities}
\label{sec:criticalF}
At criticality, the results are even more interesting.
Both the typical and average fidelities are shown in Fig.~\ref{fig:fidelity_full_phase_with_finite_size_corrections_mCan} across the phase diagram. For various system sizes, the curves exhibit a crossing near the critical point at $\delta = 0$, reminiscent of the behavior in the clean case (see Fig.~\ref{fig:clean}), where a universal value of $\sqrt{8}/\pi$ was identified for the Ising universality class~\cite{laflorencieUniversalSignaturesMajorana2023}. However, the universality class is totally different here, as any finite amount of randomness (in couplings or fields) is a relevant perturbation, and the SDRG flow drives the system to the IRFP~\cite{fisher_random_1992,fisherCriticalBehaviorRandom1995}. There, it is well-known that average critical properties look qualitatively similar to the clean case for several observables (e.g. power-law decay of bulk of surface correlations, log growth of entanglement entropy), despite their strong sample-to-sample variation and atypical distributions.  In the case of the SZM fidelity, we observe a similar trend: a finite-size crossing occurs at criticality for a nontrivial value $0<{\cal{F}}_{\rm SZM}<1$. This finite critical fidelity (numerically observed for both average and  typical values) arises from  nontrivial underlying distributions, which we discuss in detail below.

\begin{figure}
    \centering
    \includegraphics[width=1\linewidth]{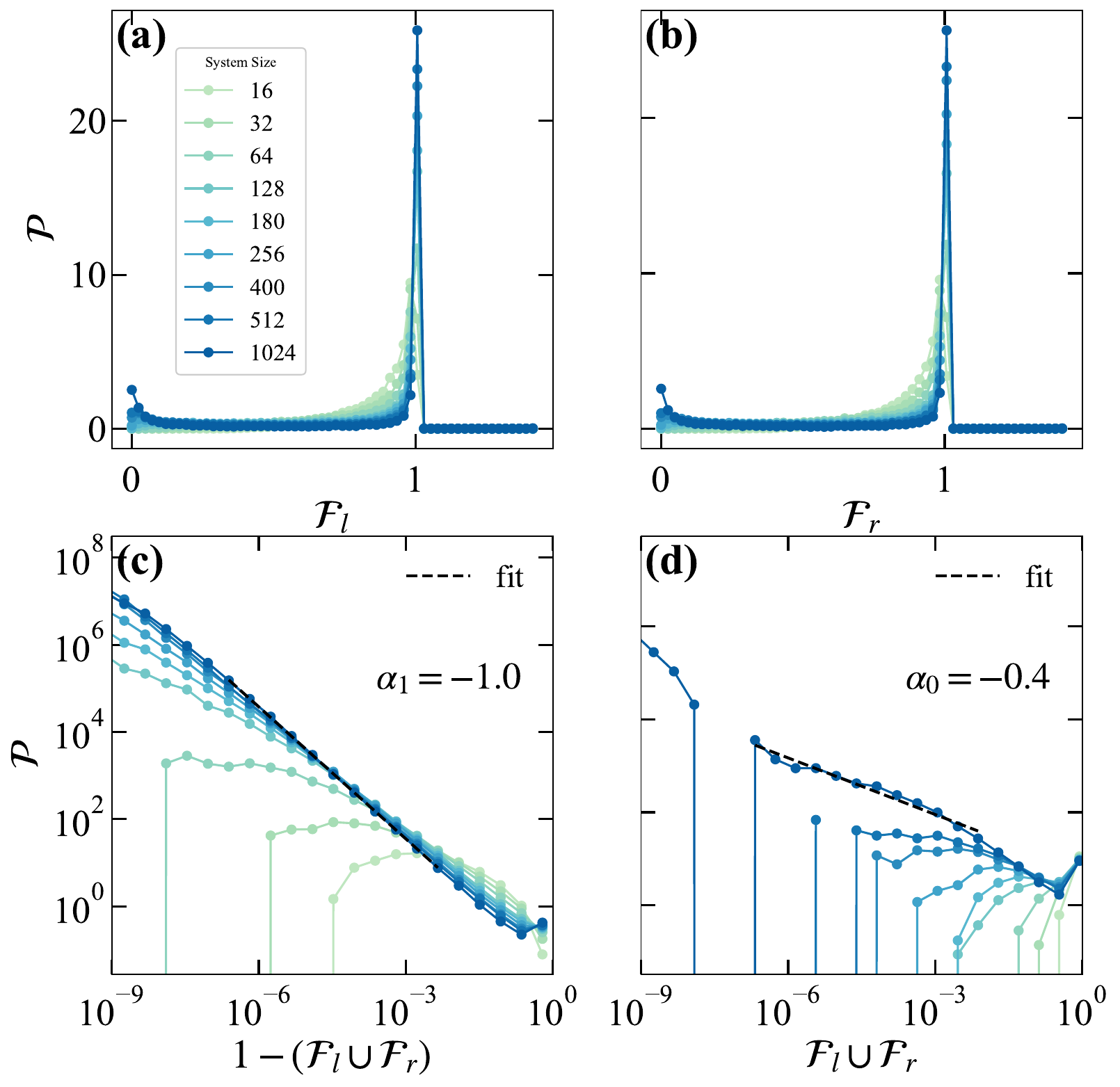}
    \caption{\label{fig:scaling_inv_fidelity_irfp_log_log} Distributions of left and right fidelities at criticality $\delta=0$ for microcanonical disorder, shown for various chain lengths $L=16,\,\ldots,\,1024$. 
    Histograms of $\mathcal{F}_l$ (a) and  $\mathcal{F}_{r}$ (b) are similar: they develop a double peak structure, more pronounced at 1.
    (c) Log-log plot of histograms of $1-(\mathcal{F}_l\cup \mathcal{F}_r)$ showing a power law behaviour for the peak at $\mathcal{F}_{l,r}=1$, with exponent $\alpha=-1$, (d) same as (c) for $(\mathcal{F}_l \cup \mathcal{F}_r)$ in the vicinity of 0. All the plots correspond to a microcanonical ensemble with disorder strength $W=3.0$ and $2\times 10^4$ samples.}
    
\end{figure}

\begin{figure}
    \centering
    \includegraphics[width=1\linewidth]{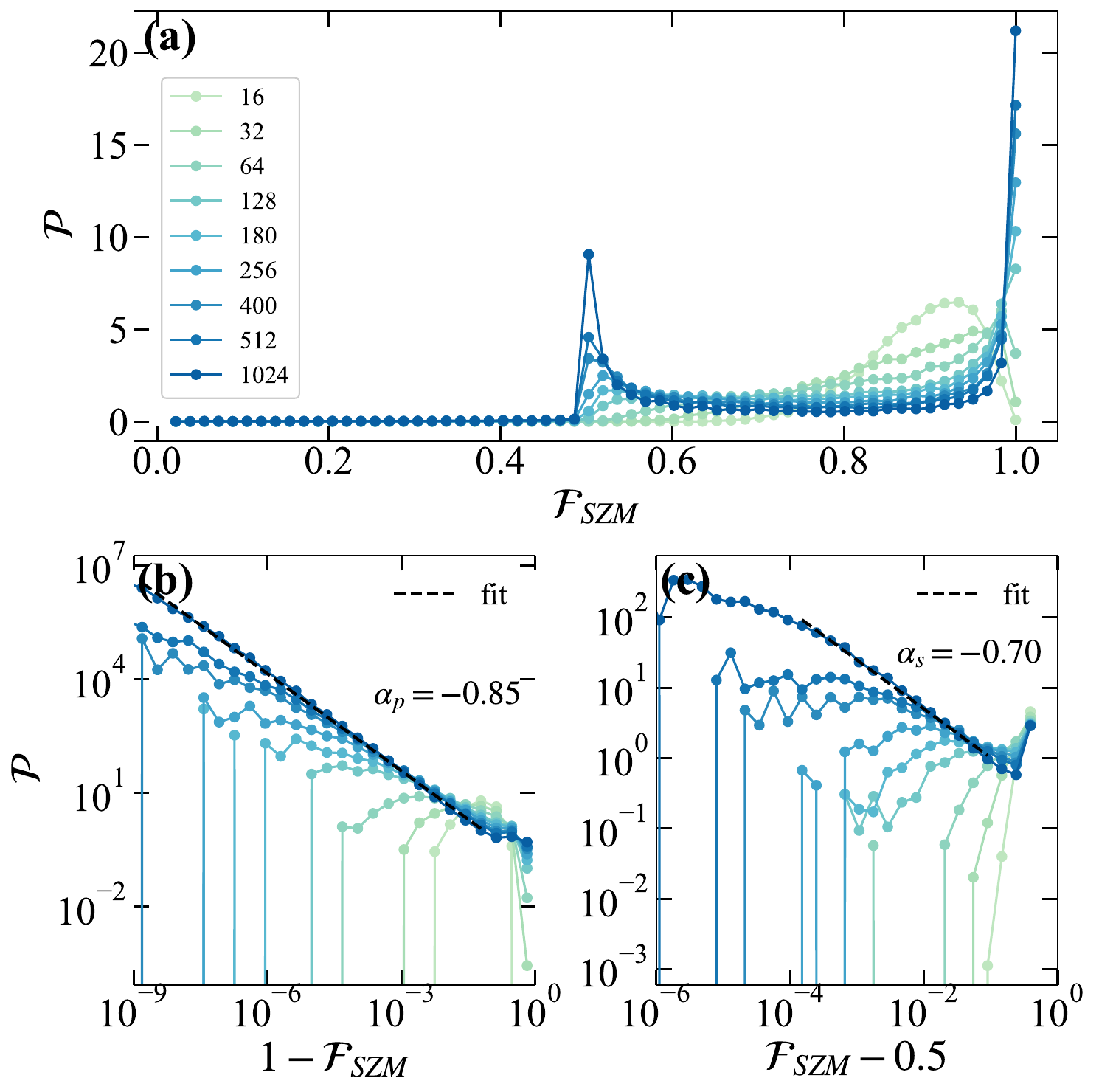}
    \caption{\label{fig:fidelity_dist_mce} (a) Distribution of the symmetrized fidelity showing a primary peak at $1$ and a secondary peak at $0.5$, (b) Logarithmic histogram of $(1-\mathcal{F}_{\rm SZM})$ indicating a power law behaviour of the primary peak with an exponent $\approx -0.85$, (d) Logarithmic histogram of the $(\mathcal{F}_{\rm SZM}-0.5)$ showing a power law behaviour with exponent $\alpha\approx0.7$. All the plots correspond to disorder strength $W=3.0$ and $2\times 10^4$ samples. }
    
\end{figure}

\subsubsection{Distribution of left and right fidelities}
Fig.~\ref{fig:scaling_inv_fidelity_irfp_log_log} shows the distributions of the left and right fidelities at the IRFP. The two distributions, ${\cal{P}}(\mathcal{F}_{l,r})$, are identical and display a bimodal, double-peaked structure, which are asymmetric, with a pronounced peak at $\mathcal{F}_{l,r} = 1$ that is significantly higher than the peak at $\mathcal{F}_{l,r} = 0$. At first sight this is surprising as one would expect that the SZM do not survive at IRFP, but the peaks in the distributions of $\mathcal{F}_l$ and $\mathcal{F}_r$ at $1$ seem to suggest that some topological features remain for a sizeable proportion of samples.
Conversely, a smaller but non-negligible density of samples, contributing to the peak at $\mathcal{F}_{l,r} = 0$,  correspond to a presumed absence of SZM.

Plotting the very same histograms in log-log scale, see panels (c-d) of Fig.~\ref{fig:scaling_inv_fidelity_irfp_log_log}, reveals power-law singularities of the two peaks at 1 and 0, following the forms
\begin{equation}
\mathcal{P}\!\bigl(\mathcal{F}_{l,r}\bigr)\sim  
  \begin{cases}
    \displaystyle \bigl(1-\mathcal{F}_{l,r}\bigr)^{\alpha_1}
      & \mathcal{F}_{l,r}\to 1\\[8pt]
    \displaystyle \bigl(\mathcal{F}_{l,r}\bigr)^{\alpha_{0}}
      & \mathcal{F}_{l,r}\to 0,
  \end{cases}
\end{equation}
with exponents $\alpha_1\approx -1$ and $\alpha_0\approx -0.4$ for the largest system size considered here $L=1024$. Note that while $\alpha_1$ seems to be reasonably converged, we observe stronger finite-size corrections for $\alpha_0$, which has not yet reached its asymptotic value. This slow crossover phenomenon will be discussed in more details below.

\subsubsection{Symmetrized fidelity at criticality}
In order to take into account both edges in each sample, the symmetrized fidelity 
\be
\mathcal{F}_{\rm SZM}=\left({\cal{F}}_{l}+{\cal{F}}_{r}\right)/2\ee
is considered.
The left and right contributions are not independent quantities, but are instead correlated, as can be seen by re-expressing Eq.~\eqref{eq:Flrdis} at the microcanonical IRFP ($\delta=0$) for each sample:
\be
    \mathcal{F}_{l} = \cfrac{m_{r}^{s}}{{\mathcal{N}}_{l}\Delta'}\quad{\rm and} \quad  \mathcal{F}_{r} = \cfrac{m_{l}^{s}}{{\mathcal{N}}_{r}\Delta'},
    \label{eq:leftandright}
\ee
$\Delta'=\cfrac{\Delta}{2 X_{\rm typ}}$ being he normalized gap.  From Eq.~\eqref{eq:leftandright}, one immediately sees that both fidelities depend on opposite edges properties. One can further derive (see App.~\ref{sec:app_fidelities}) a cross-correlation expression of the form ${\cal F}_l\sim 1/{\cal F}_r$, from which we expect that ${\cal F}_l \sim 1$ and ${\cal F}_r \sim 1$ is more likely than observing both fidelities $\approx 0$ (as numerically checked in Fig.~\ref{fig:cross_corr_leftright}).

Therefore, the distribution of symmetrized fidelity ${\cal{P}}(\mathcal{F}_{\rm SZM})$ is not a simple convolution of ${\cal P}(\mathcal{F}_{l})$ and ${\cal P}(\mathcal{F}_{r})$, which would result in a triple peak histogram with peaks at $\{0,1/2,1\}$. Instead, we observe in Fig.~\ref{fig:fidelity_dist_mce} (a) a double peak distribution, with peaks at $\mathcal{F}_{\rm SZM} = \{0.5,1\}$ separated by a valley which drifts closer to zero with increasing system size, and with a vanishing probability of $\mathcal{F}_{\rm SZM}<0.5$.

\begin{figure*}
  \centering
  \begin{minipage}[t]{0.35\textwidth}
    \centering
    \includegraphics[height=5.46cm]{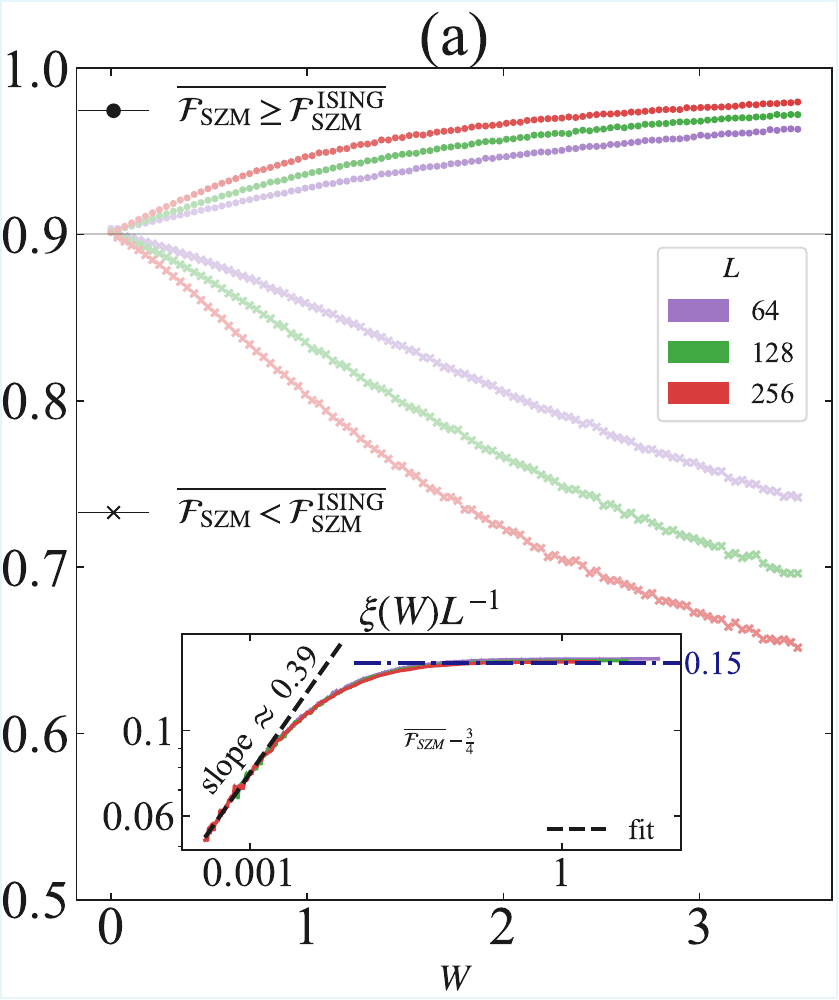}
   \label{fig: fidelity_split_flow_mce}
  \end{minipage}\hfill
  \begin{minipage}[t]{0.65\textwidth}
    \centering
    \includegraphics[height=5.7cm]{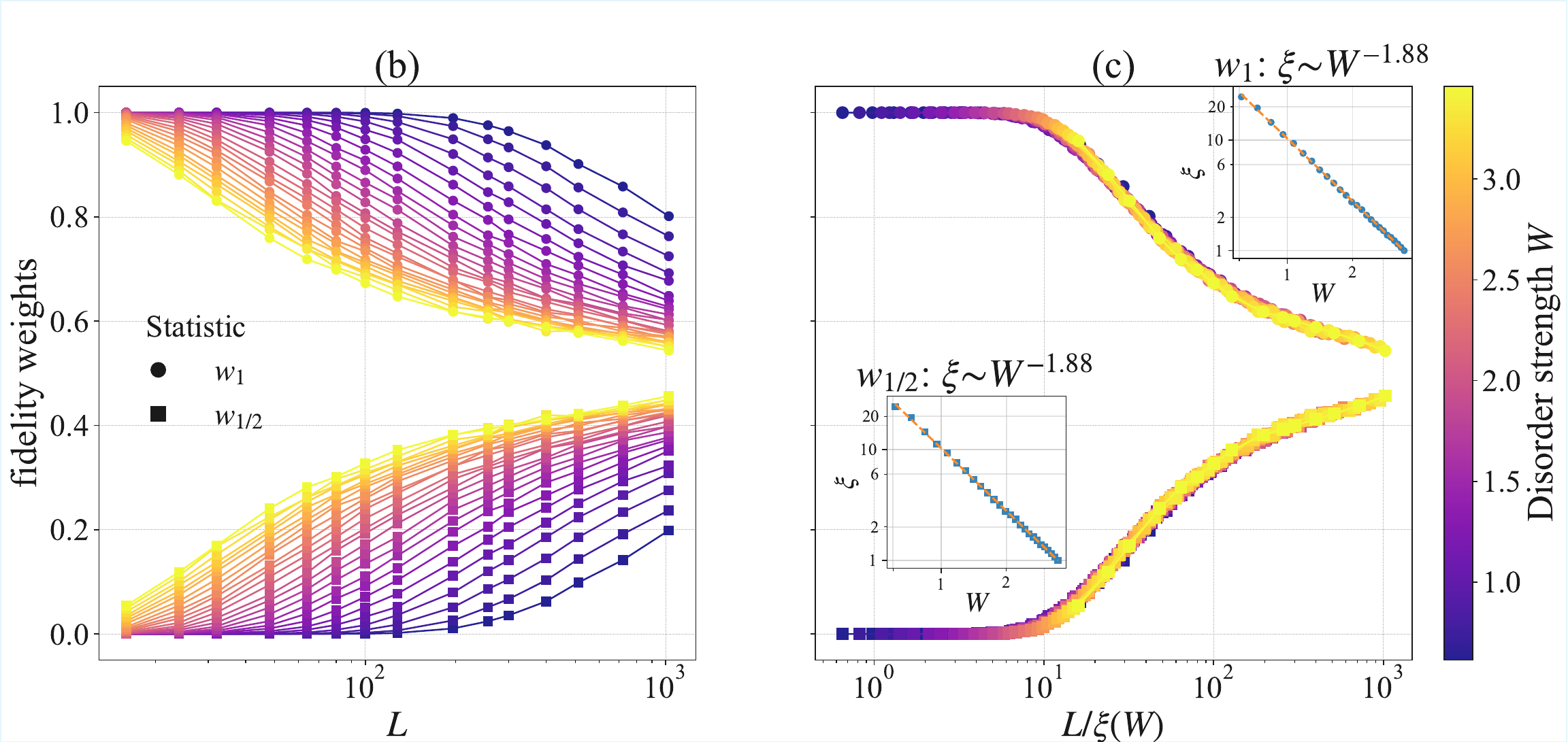}
  \end{minipage}
  \caption{Asymptotic behaviour of symmetrized fidelity at the IRFP ($\delta=0$). (a) Bi-directional flow of $\mathcal{F}_{\rm SZM}(L,W)$ starting from its clean value $\sqrt{8}/\pi$ at $W=0$ marked by the horizontal line. The curves above (below) this line, marked by $\bullet$ ($\times$) are the disorder-averaged values of $\mathcal{F}_{\rm SZM}(L,W)$ that are greater (less) than  $\mathcal{F}^{ISING}_{SZM}=\sqrt{8}/\pi$, plotted as a function of increasing disorder strength $W$. Different colors correspond to different system sizes. Color intensity (light $\rightarrow$ dark) represents the relative accumulation of $\mathcal{F}_{\rm SZM}$ values close to $1$ (upper curve) or $0.5$ (lower curve). The color gradient is normalized for each $L$ by the peak accumulation obtained across all values of $W$ for that $L$, see Appendix \ref{app: color_gradient}. Inset: Log-log plot of $({\overline{\mathcal{F}_{\rm SZM}}}-3/4)$ against the dimensionless parameter $\xi(W)/L$, where $\xi(W) \sim W^{-1.88}$ is the crossover length scale. The plot shows a collapse of curves corresponding to different system sizes, and a power law decay $({\overline{\mathcal{F}_{SZM}}}-3/4)\sim(\xi/L)^{0.39}$ at larger values of disorder strength, see Eq.~\eqref{eq: F_crossover}. The blue dot-dashed line shows the value of $(\mathcal{F}_{SZM} - 3/4)$ for the clean case which is $\approx 0.1503$ (b) Fidelity weights $w_1$ and $w_{1/2}$, see Eq.~\ref{eq: fidelity_weights}, plotted against system size for various disorder strengths $W$, shown in the colorbar. (c) Collapse of the curves in (b) obtained by single-parameter scaling. Insets: log-log plots of the scaling parameter $\xi$ {\it vs.} $W$ giving the crossover length scale $\xi\sim W^{-1.88}$. Note that this is the same crossover length scale that governs the various collapses in panels (a) and (c).\label{fig:fidelity_weight_collapse_mce}
}
\end{figure*}

This observation is particularly interesting, as it suggests that if one of the two edges yields a vanishing fidelity (for example $\bra{n,p} \Psi_r \ket{n,-p} \approx 0$) then the opposite edge instead realizes the SZM mapping ($\bra{n,p} \Psi_l \ket{n,-p} \approx 1$), as expected in the topological regime. This result supports the idea that at the IRFP, each sample (when the disorder is sampled with the microcanonical constraint) displays at least one SZM operator. 
Fig.~\ref{fig:fidelity_dist_mce} (b-c) shows the behaviour of the primary and secondary peaks at $0.5$ and $1$. The distribution of $\mathcal{F}_{\rm SZM}$ close to the peaks has the power-law form
\begin{equation}
\label{eq:F_distribution}
\mathcal{P}\!\bigl(\mathcal{F}_{l,r}\bigr)\sim  
  \begin{cases}
    \displaystyle \bigl(1-\mathcal{F}_{\rm SZM}\bigr)^{\alpha_p}
      &  \text{if } \mathcal{F}_{\mathrm{SZM}}\lesssim 1,\\[6pt]
    \displaystyle \bigl(\mathcal{F}_{\rm SZM}-1/2\bigr)^{\alpha_{s}}
      & \text{if } \mathcal{F}_{\mathrm{SZM}}\gtrsim 1/2
  \end{cases}
\end{equation}
with exponents $\alpha_p\approx -0.85$ and $\alpha_{s}\approx -0.7$, as given by the best fits performed over the largest chains $L=1024$.

\subsection{Asymptotic behavior at IRFP}
\subsubsection{Crossover of the double-peak distributions}
Even though we are dealing with quite large system sizes, Fig.~\ref{fig:fidelity_dist_mce} (a) clearly shows that the double peak distribution continues to develop upon increasing $L$. To better understand this finite-size crossover towards the infinite-randomness asymptotic behavior of the SZM fidelity, we now proceed with a systematic study upon varying both $L$ and $W$. In order to quantify the crossover of the distributions ${\cal P}({\cal F}_{\rm SZM})$ shown in Fig.~\ref{fig:fidelity_dist_mce}, we define the following weights
\begin{equation}
    \begin{split}
        & w_{1/2}(L,W)= \int_{1/2}^{3/4}d\mathcal{F}_{\rm SZM} \mathcal{P}(\mathcal{F}_{\rm SZM})\\
        & w_{1}(L,W) = \int_{3/4}^{1}d\mathcal{F}_{\rm SZM} \mathcal{P}(\mathcal{F}_{\rm SZM}),    \label{eq: fidelity_weights}
    \end{split}
\end{equation}
which measure the relative contributions of the two peaks. Fig.~\ref{fig:fidelity_weight_collapse_mce} illustrates how these contributions from the two peaks evolve. In panel (b), we observe that increasing $L$ and $W$ produces qualitatively similar effects: the asymmetry between the two contributions gradually diminishes. Specifically, $w_1$ decreases while $w_{1/2}$ increases; both slowly approaching $1/2$ (with $\sim 1/\sqrt{L}$ corrections, see App.\ref{app: szm_weight_asymptotic_mce}.
Using a one-parameter scaling, i.e. rescaling the length $L\to L/\xi(W)$, one can achieve a very good collapse of the data for both $w_1$ and $w_{1/2}$, see panel (c) for a wide range of sizes $L$ and disorder strengths $W$. The extracted crossover length scale behaves similarly for the two peaks, in both case very well described by a power-law $\xi(W)\sim W^{-\nu'}$, with $\nu'\approx 1.9$, as shown in the two insets of Fig.~\ref{fig:fidelity_weight_collapse_mce} (c). Interestingly, this disorder-dependent length scale, which governs the finite-size crossover of the SZM fidelity distributions from the clean (repulsive) fixed point to the (attractive) IRFP, closely mirrors the corresponding length scales previously identified for correlation functions~\cite{laflorencie_crossover_2004} and entanglement entropy~\cite{Laflorencie_2022}.
 
As the collapsed curves for $w_1$ and $w_{1/2}$ approach $1/2$ symmetrically, and 
given the sharply peaked singularities, we expect the probability density function 
of the symmetrized fidelity to asymptotically take a symmetric bimodal form 
$    \mathcal{P}_{\rm IRFP}(\mathcal{F}) \sim  
    {\delta(\mathcal{F}- 0.5) + \delta(\mathcal{F}- 1)}$,
yielding an average~\footnote{Given this asymptotic bimodal distribution at the IRFP, the typical symmetrized fidelity is expected to be different from the average $3/4$, with ${{\cal{F}}^{\rm typ}_{\rm SZM}}\to 1/\sqrt{2}$.}
 fidelity 
\be
{\overline{{\cal{F}}^{\rm IRFP}_{\rm SZM}}}\to \frac{3}{4}.\ee

\subsubsection{Crossover of the SZM fidelity from clean to IRFP}
The crossover towards the IRFP asymptotic result can be further checked for the average SZM fidelity, as shown in Fig.~\ref{fig:fidelity_weight_collapse_mce} (a) where we plot three quantities. In the main panel, we show the two averages computed over ${\cal{P}}({\cal F}_{\rm SZM})$ (e.g. from Fig.~\ref{fig:fidelity_dist_mce}) for the right and left peaks, which respectively go towards 1 and 1/2 when $W$ and/or $L$ are increased. Perhaps more interesting, the inset displays a data collapse for the full average, which takes the following scaling form
\be
\label{eq: F_crossover}
{\overline{{\cal{F}}^{\rm IRFP}_{\rm SZM}}}-\frac{3}{4}\to  
  \begin{cases}
    \displaystyle \left(\frac{\xi}{L}\right)^{-a}\quad
      & {\rm if}\quad L\gg\xi\\[12pt]
    \displaystyle  \frac{\sqrt 8}{\pi}-\frac{3}{4}\approx 0.1503
      & {\rm if}\quad L\ll\xi,
  \end{cases}
\ee
using the {\it{same}} disorder-dependent length scale $\xi=\xi(W)$ as the one used to obtain the collapse of the weights in Fig.~\ref{fig:fidelity_weight_collapse_mce} (c). The asymptotic IRFP value is numerically found to be attained algebraically $\sim (\xi/L)^{-a}$, with an exponent $a\approx 0.39$. 

\section{Discussions}
\label{sec:disc}
\subsection{Sampling the disorder: ensemble dependence}
So far we have mostly discussed the microcanonical disorder, which enforces each sample to have the exact constraint
\be 
\frac{1}{L-1}\sum_{j=1}^{L-1}\ln X_i-\frac{1}{L}\sum_{j=1}^{L}\ln h_i=\delta,
\ee
while canonical sampling (see App.~\ref{app:can}) allows for finite-size fluctuations of the control parameter $\delta$, which,  from central limit theorem typically goes like $\sim 1/\sqrt{L}$. Nevertheless, regardless of whether the global constraint is imposed, the canonical and microcanonical ensembles display qualitatively similar behavior for both the average and typical SZM fidelities off-criticality ($\delta \neq 0$).

At the IRFP, however, the canonical ensemble imposes fewer constraints at the level of individual samples, which has significant consequences for the correlations between the surface magnetizations at the two edges, as discussed by Dhar and Young~\cite{dharEnsembleDependenceRandom2003} and Monthus~\cite{monthusFinitesizeScalingProperties2004a}.
Hence, the critical distributions of the SZM fidelities (see Fig.~\ref{fig:scaling_inv_fidelity_irfp_log_log_can} and Fig.~\ref{fig:fidelity_dist_ce}), instead of taking the bimodal form $\{1/2,\,1\}$ observed in the microcanonical case, display a trimodal shape $\{0,\,1/2,\,1\}$. This result can be interpreted as follows. Whereas in the microcanonical case each sample was found to support at least one (and possibly two) SZM operators, canonical sampling also allows for realizations with no SZM, which is typical of the trivial regime. However, inspection of the numerical histograms in Fig.~\ref{fig:fidelity_dist_ce} does not readily allow one to anticipate the asymptotic canonical distribution, particularly whether samples with vanishing SZM fidelity persist in the thermodynamic limit, where canonical and microcanonical ensembles are expected to become equivalent.

\subsection{Connections with average duality and anomalies}
We can also try to connect our results to recent developments on anomalies and average symmetries in one-dimensional random systems~\cite{li_average_2026,panahi_quantum_2026,ma_topological_2025,ma_average_2023}. In particular, average non-invertible categorical symmetries, such as the Kramers–Wannier (KW) duality, leads in disordered Ising-Majorana chains~\cite{li_average_2026} to power-law decaying average correlations~\cite{panahi_quantum_2026} and long-range entanglement~\cite{li_average_2026}.

In our setting—more specifically, for microcanonical sampling of the disorder—one may ask whether the observed resilience of SZMs at one or both edges reflects a complementary structure: when one edge fails to provide a high-fidelity mapping between parity sectors, the opposite edge compensates. This statistical complementarity suggests that, for each sample, the microcanonical constraint induces an edge-selective structure that shows up in the symmetrized SZM fidelity. This could also be interpreted as an additional microscopic manifestation of the average KW duality.

\subsection{Disordered Rydberg atom arrays}

A natural arena to make contact with experiments is provided by disordered arrays of Rydberg atoms, which have recently emerged as a versatile platform for realizing constrained spin chains with tunable disorder \cite{prodius_interplay_2026,soto_infinite_2026,li_random_2025,yue_observation_2025,brodoloni_spin-glass_2025}. Interestingly, the interplay of localization, topology/dimerization has been explored in such systems, as well as the possible emergence of infinite-randomness physics.

Certain Rydberg chains can be effectively described by disordered Ising, Kitaev-type, or SSH models, where edge modes and domain-wall excitations dominate the dynamics. Using a fidelity-based characterization, we showed that SZMs exhibit some edge robustness, leading to measurable asymmetries in edge responses under disorder. The complementarity of left and right edge fidelities can be detected via boundary-sensitive probes, including quench dynamics of edge spins.

It would be fascinating to leverage the local control available in Rydberg arrays to engineer regimes in which specific disorder profiles effectively support high-fidelity SZMs on one or two edges. This could establish a quantitative link between microscopic operator diagnostics and experimentally accessible boundary observables. In this context, it is also worth highlighting the recent, remarkable observation of Majorana edge physics using superconducting qubit platforms~\cite{MiSonnerNiuEtAl2022,jin_topological_2025}, a problem which has also been theoretically investigated  in the framework of random-Field Floquet quantum Ising model~\cite{schmid_robust_2024,mockel_floquet_2026}.

\section{Conclusions}
\label{sec:conclusion}
In this work, we have investigated the robustness of strong zero modes 
(SZMs) in the random Ising-Majorana chain using the SZM fidelity 
$\mathcal{F}_{\rm SZM}$ as a many-body topological diagnostic that probes the 
faithfulness of the parity-sector mapping across the {\it{entire}} 
many-body spectrum.

In the topological phase ($\delta > 0$), we have shown that SZMs are 
robust to the presence of quenched disorder, with fidelities converging 
exponentially to unity, $1 - \mathcal{F}_{\rm SZM} \sim \exp(-L/\xi)$, 
even deep in the Griffiths regime where the bulk gap has closed. 
The associated length scale diverges at criticality as 
$\xi \sim |\delta|^{-\nu}$, with $\nu \approx 2$ for both average and 
typical fidelities, consistent with the {\it{average}} correlation length exponent of the 
infinite-randomness fixed point (IRFP). In the trivial phase ($\delta < 0$), 
the average fidelity vanishes as a power law, $\mathcal{F}_{\rm SZM} \sim 
L^{-\omega}$ with $\omega \approx 1.2$, while the typical fidelity decays 
exponentially.\\

The most striking results emerge precisely at the IRFP ($\delta = 0$), 
where the fidelity distributions develop a singular structure that is 
strongly ensemble dependent. In the {\it{microcanonical}} ensemble, the 
individual left and right fidelities display a bimodal distribution with 
power-law peaks at $\{0, 1\}$, while the symmetrized fidelity $\mathcal{F}_{\rm SZM}$ 
develops a double-peak structure at $\{1/2, 1\}$ with a vanishing 
weight below $1/2$. This edge-selective complementarity, where a 
vanishing fidelity at one edge is ``compensated'' by a near-unit fidelity at 
the other, implies that each microcanonical sample supports at least one 
SZM operator, a remarkable manifestation of localization-protected 
topological order that seems to persist at criticality. In the asymptotic regime, the two peaks 
carry equal weight, yielding a fixed-point critical fidelity
$\mathcal{F}_{\rm SZM}^{\rm IRFP} \to \frac{3}{4}$.

In the {\it{canonical}} ensemble, the relaxation of the per-sample 
constraint allows for disorder realizations that fall into the trivial 
regime, generating a qualitatively different triple-peak structure 
$\{0, 1/2, 1\}$ for $\mathcal{F}_{\rm SZM}$. The presence of a peak at 
zero signals a finite fraction of samples with no topological character, 
in sharp contrast to the microcanonical case. We nevertheless expect this to be finite-size physics, as  both ensembles should become equivalent in the thermodynamic limit.\\

We have also argued that the edge-selective structure of the microcanonical 
critical distributions could be a microscopic boundary manifestation of the 
average Kramers-Wannier (KW) duality symmetry at the IRFP: the SZM and its 
dual (a nonlocal string operator) are placed on equal footing by the 
microcanonical constraint, which fixes every sample to the self-dual 
point $\delta = 0$ exactly. Our results thus establish $\mathcal{F}_{\rm SZM}$ 
as a sharp, operationally meaningful probe of localization-protected 
topological order, capable of distinguishing clean and infinite-randomness 
universality classes, and of revealing the interplay between disorder 
ensembles, boundary topology, and emergent average symmetries in 
one-dimensional random quantum systems.\\

{\it{Note added:}} While completing this work, we came across a very recent posting by Moudgalya and Motrunich~\cite{moudgalya_strong_2026}, which develops a systematic algebraic framework for constructing SZMs, demonstrates that integrability is not required for their existence, and identifies two fundamentally distinct classes of SZMs. Extending their analysis to the random case would be particularly interesting.

\acknowledgments
We thank J. Kemp and C. Monthus for their critical reading of the manuscript. We also acknowledge N. Chepiga for collaborations on related topics. This work has been partly supported by the ANR research grant ManyBodyNet No. ANR-24- CE30- 5851. We acknowledge the use of HPC resources from CALMIP (grants 2025-P0677).\\

\noindent
{\bf Data availability:} All data supporting this work are available upon request.\\

\appendix
%%%%%%%%%%%%%%%%%%%%%%%
\setcounter{section}{0}
\setcounter{secnumdepth}{3}
\setcounter{figure}{0}
\setcounter{equation}{0}
\setcounter{table}{0}
\renewcommand\thesection{A\arabic{section}}
\renewcommand\thefigure{A\arabic{figure}}
\renewcommand\theequation{A\arabic{equation}}
\renewcommand\thetable{A\arabic{table}}

\renewcommand{\theHsection}{A\arabic{section}}
\renewcommand{\theHfigure}{A\arabic{figure}}
\renewcommand{\theHtable}{A\arabic{table}}
\renewcommand{\theHequation}{A\arabic{equation}}

\onecolumngrid
\section{Free fermion calculations}
\subsection{Jordan Wigner transformation}
\label{sec:app_jw}
Jordan-Wigner transformation maps spin (Pauli) operators into fermionic operators and vice-versa. Using these we can transform a spin Hamiltonian to a fermionic Hamiltonian or Majorana hamiltonian and vice versa. The transformations are
\begin{align}
    &\sigma^x_j = K_j \big(c^{\dagger}_j + c_j\big)\\
    &\sigma^y_j = iK_j \big(c^{\dagger}_j - c_j\big)\\
    &\sigma^z_j = 1-2c^{\dagger}_j c_j\\
    &\text{where}\quad  K_j=\prod_{k=1}^{j-1}\sigma^z_k
\end{align}
\subsection{Diagonalization of the Kitaev chain}
\label{sec:app_diago}
In the random case, we use the Nambu formalism~\cite{youngNumericalStudyRandom1996} to solve the inhomogeneous free-fermion (Kitaev chain) Hamiltonian Eq.~\eqref{eq:Kitaev_hamiltonian}.
Introducing $\varphi^{\dagger}=\left(c_1^\dagger\cdots c_L^\dagger\,c_1\cdots c_L\right)$, we simply get a matrix representation of the Hamiltonian:
\begin{equation}
\label{eq:Hmat}
{\cal{H}}_{K}= \varphi^\dagger
\begin{pmatrix}
\cal M&{\widetilde{\cal M}}\\
-{\widetilde{\cal M}} &-\cal M\\
\end{pmatrix}\varphi.
\end{equation}
$\cal M$ and ${\widetilde{\cal M}}$ are $L\times L$ matrices that have the following properties.
\be
{\cal M}_{j,j}=\frac{\mu_j}{2}=h_j,\quad 
{\cal M}_{j,j+1}={\cal M}_{j+1,j}=-\frac{t_j}{2}=-\frac{X_j+Y_j}{4},
\ee
while ${\cal M}_{i,j}=0$ elsewhere (unless periodic boundary conditions are used).
%: $M_{1,L}=M_{L,1}=p{t_L}/{2}$, where $p=\pm 1$ is the fermionic parity).
The matrix ${\widetilde{\cal M}}$ is anti-symmetric: 
\be
{\widetilde{\cal M}}_{j,j+1}=-{\widetilde{\cal M}}_{j+1,j}=-\frac{\Delta_j}{2}=-\frac{X_j-Y_j}{4},
\ee
and ${\widetilde{\cal M}}_{i,j}=0 \,\,{\rm{otherwise}}$.
This free-fermion problem can be solved by diagonalizing the $2L\times 2L$ matrix Eq.~\eqref{eq:Hmat}, the quadratic Kitaev Hamiltonian is then rewritten as
\be
{\cal{H}}_{K}=2\sum_{m=1}^{L}\epsilon_m\left(\phi^{\dagger}_m\phi_{m}^{\vphantom{\dagger}}-\frac{1}{2}\right),
\label{eq:Hdiag}
\ee
with single particle energies $0\le \epsilon_1\le\epsilon_2\le\ldots \epsilon_L$, and new fermionic modes that are given by the Bogoliubov transformation 
\be
\phi_m^{\dagger}=\sum_{j=1}^{L}\left(u^m_j\,c_j^\dagger+v^m_j\,c_{j}^{\vphantom{\dagger}}\right),
\ee
with real $u^m_j$ and $v^m_j$. The inverse transformation is
\be 
c_{j}^{{\dagger}}=\sum_{m=1}^{L}\left(u_j^m \phi_{m}^{{\dagger}}+v_j^m \phi_{m}^{\vphantom{\dagger}}\right).
\label{eq:inversetransformation}
\ee

\subsection{Strong zero mode operators and fidelities}
\label{sec:app_szm}
For Eq.~\eqref{eq:inversetransformation}, one can  express the SZM operators Eq.~\eqref{eq:SZM_lr} as follows
\bea
\Psi_l &=& \frac{1}{{\cal{N}}_l}\sum_{j=1}^{L} \Theta_{j-1}^{a}a_j=\sum_{m=1}^{L}{\cal{A}}_m\left(\phi^{\dagger}_{m}+\phi^{\vphantom{\dagger}}_{m}\right)\\
\Psi_r &=& \frac{1}{{\cal{N}}_r}\sum_{j=1}^{L} \Theta_{j-1}^{b} b_{L+1-j}=i\sum_{m=1}^{L}{\cal{B}}_m\left(\phi^{\dagger}_{m}-\phi^{\vphantom{\dagger}}_{m}\right),
\label{eq:}
\eea
with coefficients  $\Theta^a_{j>0} = \prod_{i=1}^j \cfrac{h_i}{X_i}$, $\Theta^b_{j>0} = \prod_{i=1}^j \cfrac{h_{L-i+1}}{X_{L-i}}$, and $\Theta^{a,b}_{0}=1$. 
Normalization factors ${\cal{N}}_{l,r}$ follow Eq.~\eqref{eq:norms}, and 
\be
{\cal{A}}_m =\frac{1}{{\cal N}_l}\sum_{j=1}^{L}\Theta_{j-1}^a \left(u^m_j+v^m_j\right)\quad 
{\rm{and}}\quad {\cal{B}}_m = \frac{1}{{\cal{N}}_r}\sum_{j=1}^{L}\Theta_{j-1}^b \left(u^m_{L+1-j}-v^m_{L+1-j}\right).
\ee
In the ordered regime, many-body eigenpairs only differ by their occupation of the lowest mode $m=1$, such that
\be
{\cal{F}}_{l}=
{{\cal{A}}_1\quad {\rm and}\quad {\cal{F}}_{r}={\cal{B}}_1}, 
\ee
which are clearly independent on the energy level $\ket{n,p}$.
% \subsection{Parity gap}
% \label{app:paritygap}

\subsection{Cross-correlations between left and right fidelities}
\label{sec:app_fidelities}
\begin{figure}[h!]
  \centering
   \includegraphics[width=.5\columnwidth]{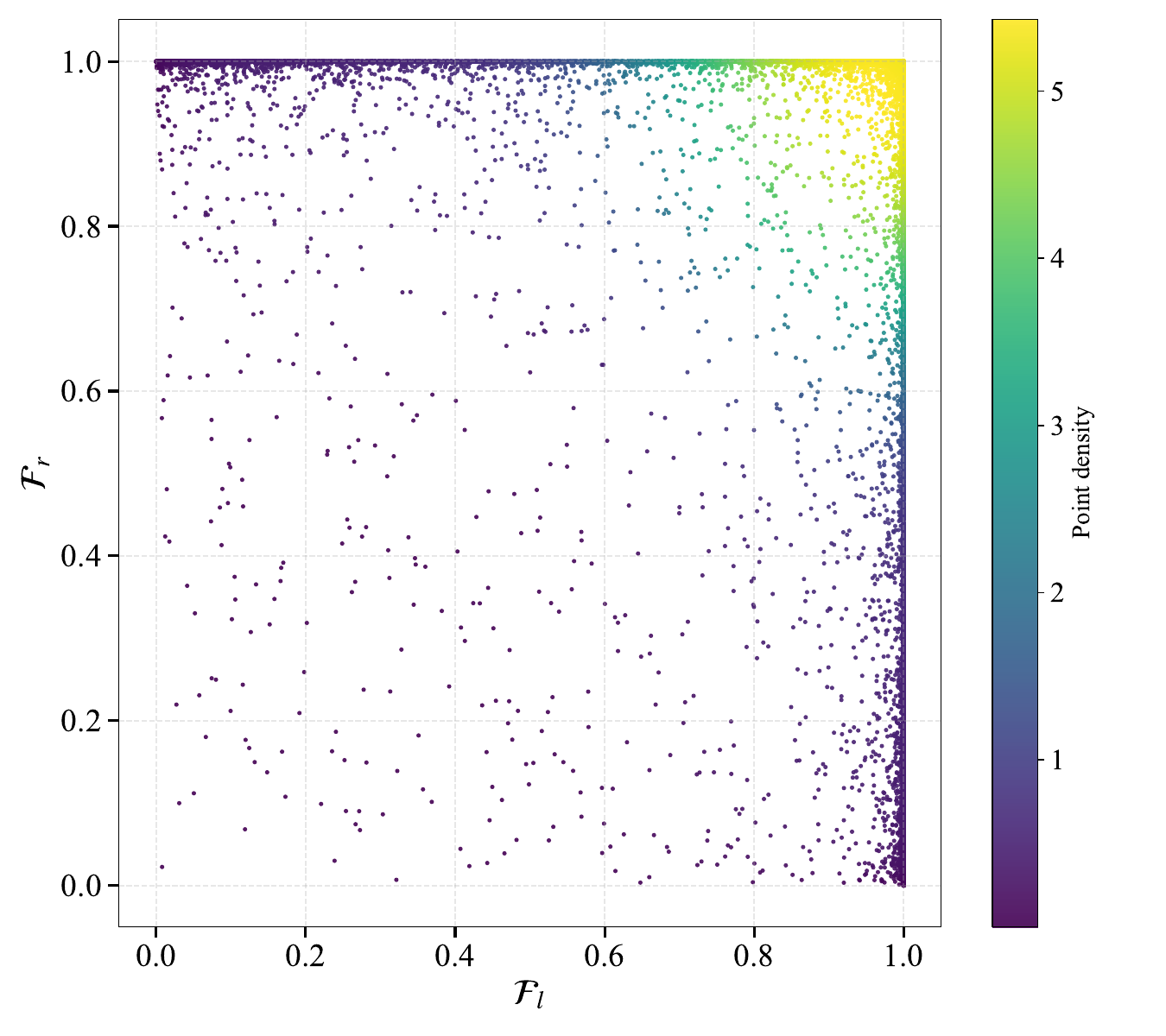}
  \caption{\label{fig:cross_corr_leftright} $(\mathcal{F}_l, \mathcal{F}_r)$ points at $\delta = 0$ for the microcanonical ensemble for a system size $L=512$. The color in any region indicates the density of points $(\mathcal{F}_l, \mathcal{F}_r)$ in that region. The total number of samples taken are $2\times 10^4$ at a disorder strength $W=2.0$.}
\end{figure}

Applying a SZM to the GS (the Bogoliubov vacuum $\ket 0$) yields 
\be
\Psi_l\ket{0}={\cal{F}}_l\ket{1}+\sum_{m>1}{\cal{A}}_m\ket{m}.
\ee
We then apply the boundary magnetization operator, and project onto the GS 
\be
\bra{0}\sigma_1^x\Psi_l\ket{0}=\frac{1}{{\cal{N}}_l}=m_l^s{\cal F}_l+\sum_{m>1}{\cal A}_m\left(u_1^m+v_1^m\right),
\ee
which after a few steps gives the following expressions for the surface magnetizations $m_{l,r}^s=\bra{0}\sigma_{1,L}^x\ket{1}$
\be
m_{l,r}^s=\frac{1-\eta_{l,r}}{{\cal N}_{l,r}{\cal F}_{l,r}},
\label{eq:surface_mlr}
\ee
with
\bea
\eta_{l}&=&\sum_{j=1}^{L}\Theta_{j-1}^{a}\sum_{m>1}\left(u_{1}^m+v_1^m\right)\left(u_j^m+v_j^m\right)\\
\eta_{r}&=&\sum_{j=1}^{L}\Theta_{j-1}^{b}\sum_{m>1}\left(u_{L}^m-v_L^m\right)\left(u_{L+1-j}^m+v_{L+1-j}^m\right).
\eea
Interestingly, combining Eq.~\eqref{eq:surface_mlr} with Eq.~\eqref{eq:leftandright} leads to
\be
{\cal F}_l {\cal F}_r=\frac{1-\eta_l}{\Delta'{\cal N}_l{\cal N}_r}=\frac{1-\eta_r}{\Delta'{\cal N}_l{\cal N}_r},
\ee
which implies that $\eta_l=\eta_r=\eta$. We then can express the cross-correlation between left and right fidelities as
\be 
{\cal F}_l =\frac{1-\eta}{\Delta'{\cal N}_l{\cal N}_r}\times \frac{1}{{\cal F}_r},
\ee 
from which we can anticipate that ${\cal F}_l \sim 1$ and ${\cal F}_r \sim 1$ is more likely than observing both fidelities being $\approx 0$, as clearly shown in Fig.~\ref{fig:cross_corr_leftright}.

\subsection{Asymptotic values of fidelity weights $w_1$ and $w_{1/2}$}
\label{app: szm_weight_asymptotic_mce}
\begin{figure}[h]
    \centering
    \includegraphics[width=0.4\linewidth]{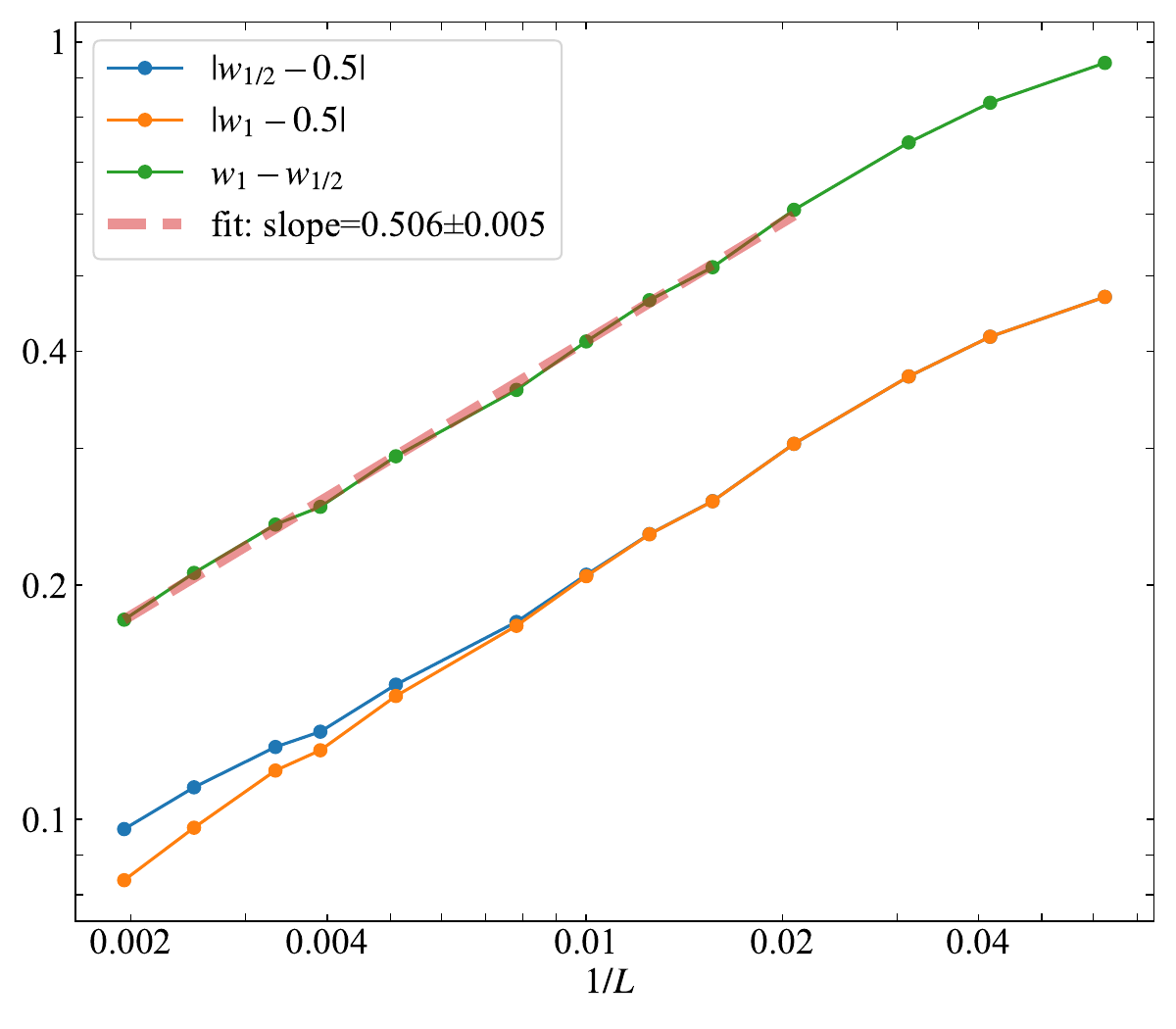}
    \caption{Log-log plot of disorder-averaged Fidelity weights $|w_{1}-0.5|$ and $|w_{1/2}-0.5|$ and $w_1 - w_{1/2}$ at IRFP for a disorder strength $W=3.0$ vs $1/L$ showing their power law decays with system size. Note that all three plots are nearly parallel to each other. The dotted red lines show a straight line fit for the $w_1-w_{1/2}$ plot. The averaging is performed for $2\times 10^4$ samples from the microcanonical ensemble.} 
    \label{fig: szm_weight_asymptotic_mce}
\end{figure}

In figure \ref{fig: szm_weight_asymptotic_mce}, we show the asymptotic behaviour of fidelity weights $w_1$ and $w_{1/2}$, defined in equation \ref{eq: fidelity_weights} in the large system size limit.
We see that $(w_1 - w_{1/2})$ shows a power law decay to zero with finite size corrections of the order $\mathcal{O}(L^{-1/2})$. Since $w_1 + w_{1/2} = 1$, it is easy to conclude that both $w_1$ and $w_{1/2}$ at large system sizes approach the value $1/2$ with finite size corrections of the order $\mathcal{O}(L^{-1/2})$. This conclusion is further strengthened by looking at the behaviour of $|w_1-0.5|$ and $|w_{1/2}-0.5|$ separately, both of which show power law decay parallel to that of $(w_1 - w_{1/2})$.

\section{Canonical Ensemble}
\label{app:can}
We repeat the computations of the main text for disordered samples belonging to the canonical ensemble. The bond strength parameter $\{X_j\}$ and the transverse field $\{h_j\}$ are both uniformly distributed such that 
\begin{equation}
    \begin{split}
        & \overline{\ln X}=1\\
        & \overline{\ln h}=1-\delta
    \end{split}
\end{equation}
\begin{figure*}[b]
  \centering
   \includegraphics[width=.8\textwidth, trim=0cm 0 0cm 0, clip]
    {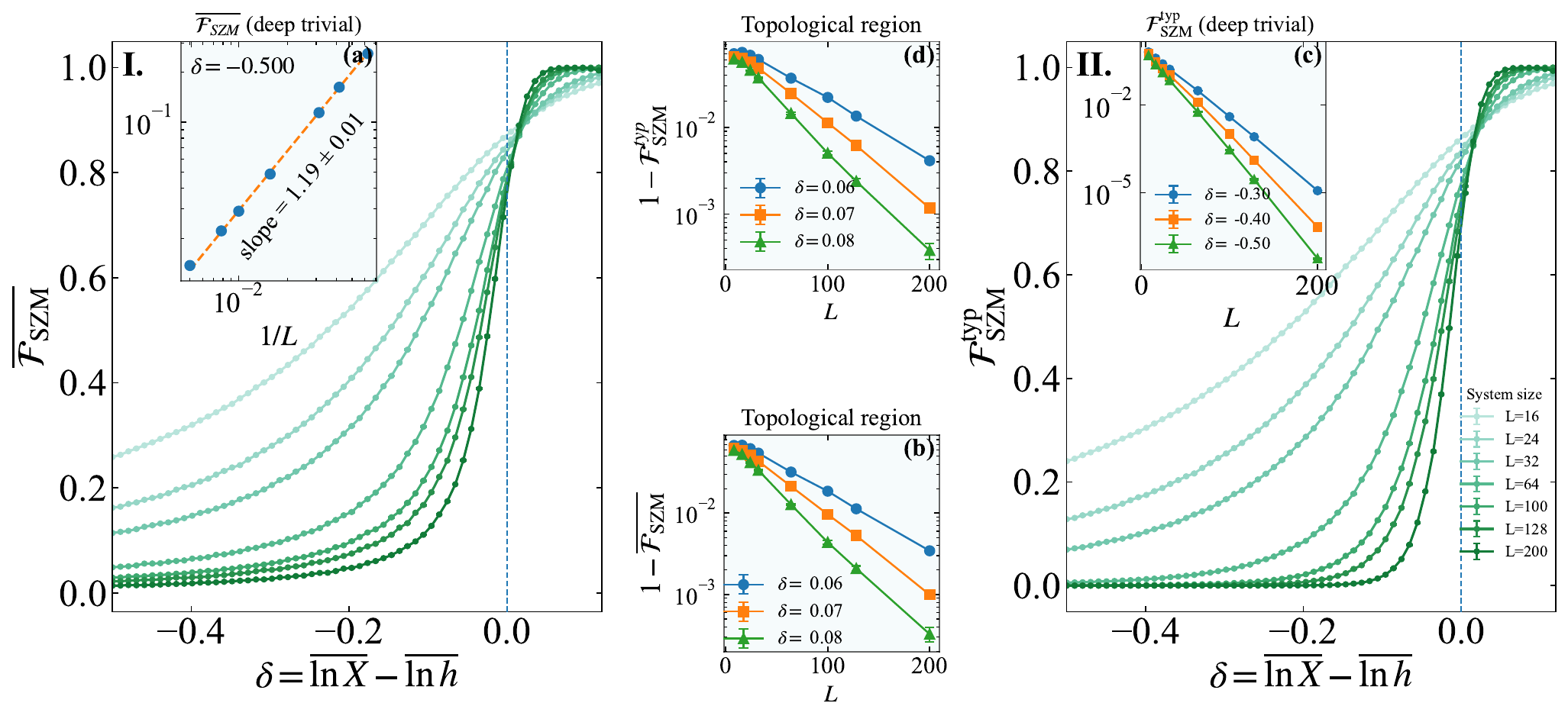}
  \caption{\label{fig:fidelity_full_phase_with_finite_size_corrections_can}Plot I: Canonical Ensemble: Disorder averaged symmetrized fidelity plotted against the control parameter $\delta$. The vertical dotted line marks the quantum critical point. Trivial and topological regions correspond to $\delta<0$ and $\delta>0$ respectively. Inset (a) : A log-log plot of $\overline{\mathcal{F}_{\rm SZM}}$ vs inverse system size, plotted at the control parameter value $\delta=-0.5$, in the deep trivial region, showing a power law decay to zero. Inset (b) : $1-\overline{\mathcal{F}_{\rm SZM}}$ plotted against $L$ on a semi-log plot at various points in the topological region, showing exponential corrections to $\overline{\mathcal{F}_{\rm SZM}}$ as it approaches $1$.
  Plot II: Same as plot I, for typical values of the symmetrized fidelity. Inset (c) Semi-log plot of $\mathcal{F}^{typ}_{\rm SZM}$ at various points inside the deep trivial region, showing a exponential decay to zero. Inset (d): Semi-log plot of $1-\mathcal{F}^{typ}_{\rm SZM}$ at various points in the topological region, showing exponential finite size corrections. The error bars, where not visible, are smaller than the markers.}
  
\end{figure*}
is obeyed at the level of the ensemble and not at the level of the sample, which means that the value of the parameter $\delta$ fluctuates as $\sim 1/\sqrt{L}$ for a sample with $L$ sites. Unlike the case for the microcanonical ensemble, neither $\{X_j\}$ nor $\{h_j\}$ are correlated. The results that follow show a strong dependence on the choice of ensemble for the disordered system, particularly at the infinite randomness fixed point.
\begin{figure*}[h!]
    \centering
    \includegraphics[width=.7\linewidth]{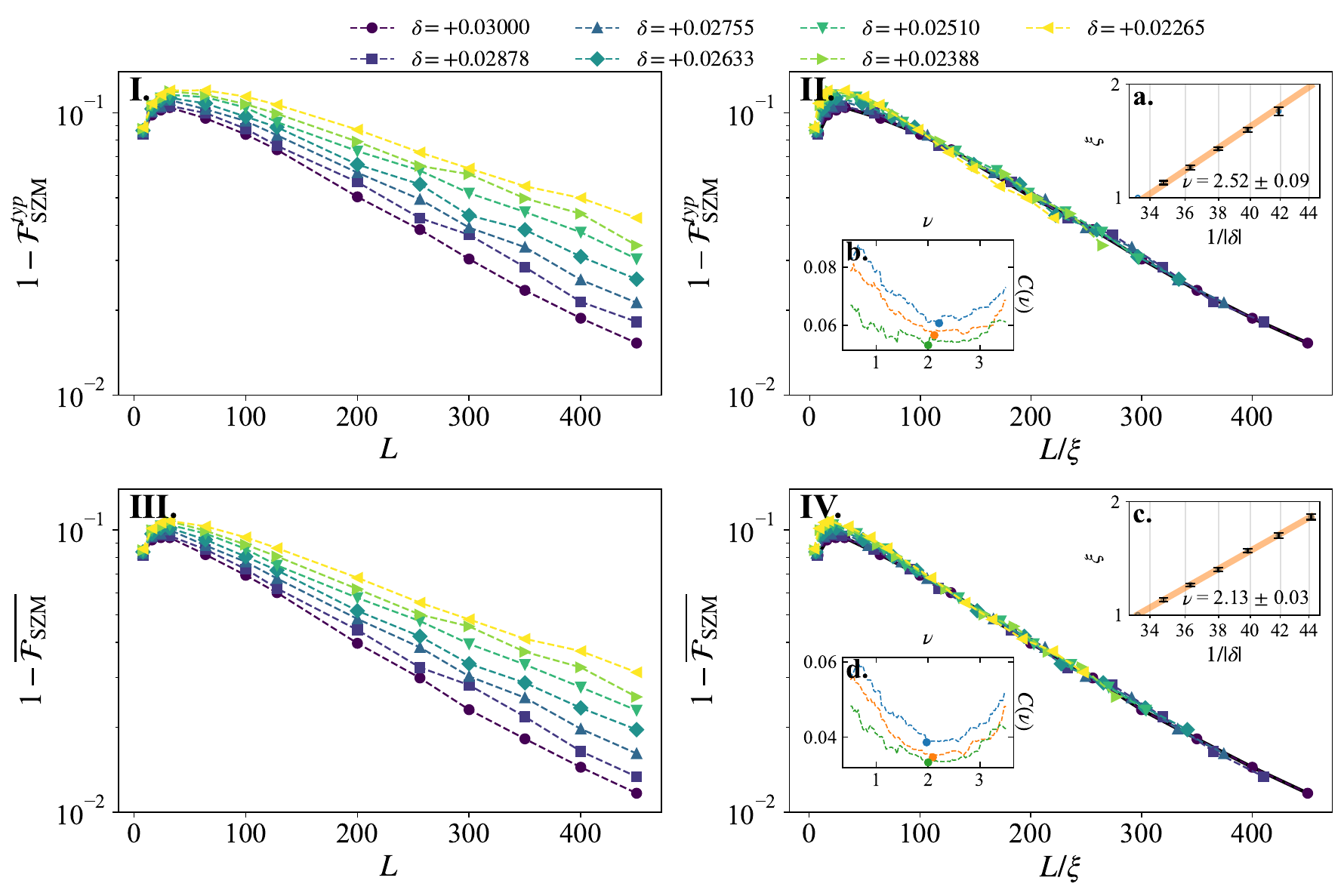}\hfill
    
    \caption{\label{fig:critical_exponents_canonical_topo}
    Canonical Ensemble: I.Approach of $\mathcal{F}_{\rm SZM}^{typ}$ to $1$ in the topological regime, close to criticality, for various values of $\delta$ on a semi-log plot. II. Collapse of the curves in I using one parameter scaling $L\rightarrow L/\xi$.  Inset (a) Length scale $\xi$ vs $1/|\delta|$ on a log-log plot showing a dependence $\xi \sim |\delta|^{-\nu}$. Inset (b)Values of the cost function $C(\nu)$ plotted against the critical exponent $\nu$. The different colors correspond to number of bins used to calculate $C(\nu)$ and the solid circles mark the minimum of the $C(\nu)$ curve. Both the insets show the value of $\nu\sim 2$.
    Figures III, IV, and insets (c) and (d) are the corresponding quantities for $\overline{\mathcal{F}_{\rm SZM}}$, which also shows a critical exponent close to $2$.
    }
    
\end{figure*}

\begin{figure}[h!]
    \centering
    \includegraphics[width=.4\linewidth]{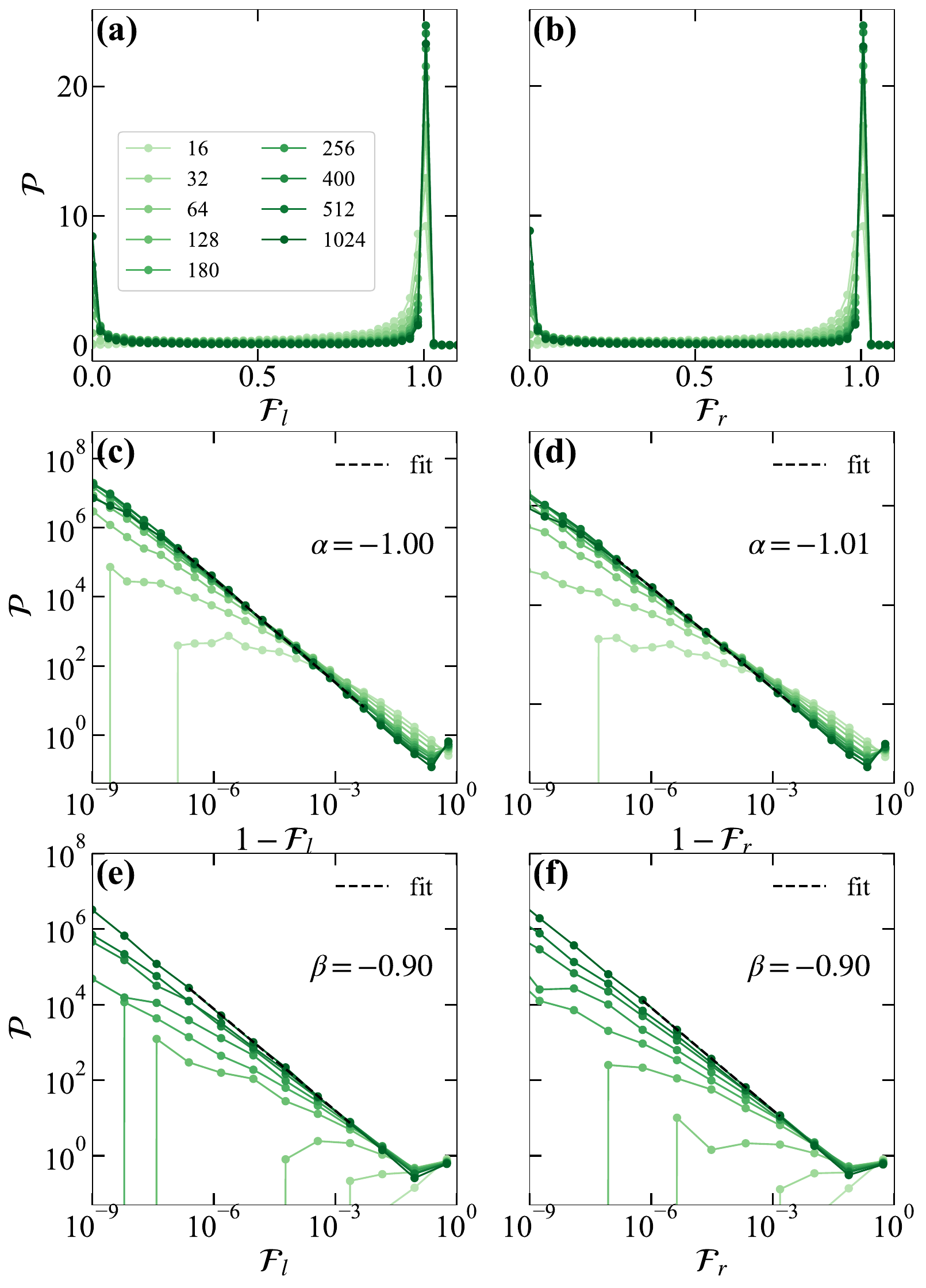}
    \caption{\label{fig:scaling_inv_fidelity_irfp_log_log_can}(a) Distribution of $\mathcal{F}_l$, (b)Distribution of $\mathcal{F}_{r}$, (c)Distribution of $(1-\mathcal{F}_l)$ on a logarithmic histogram showing a power law behaviour for the peak at $\mathcal{F}_l=1$, with exponent $\alpha=-1$, (d) same as (c) for $\mathcal{F}_r$. (e)Distribution of $\mathcal{F}_l$ on a logarithmic histogram showing a power law behaviour for the secondary peak at $\mathcal{F}_l=0$, with exponent $\alpha=-0.9$, (f) same plot as (e) for $\mathcal{F}_r$. All the plots correspond to a canonical ensemble with disorder strength $W=3.0$. The number of samples used is $2\times 10^4$ }
\end{figure}

\begin{figure}[h!]
    \centering
    \includegraphics[width=.6\linewidth]{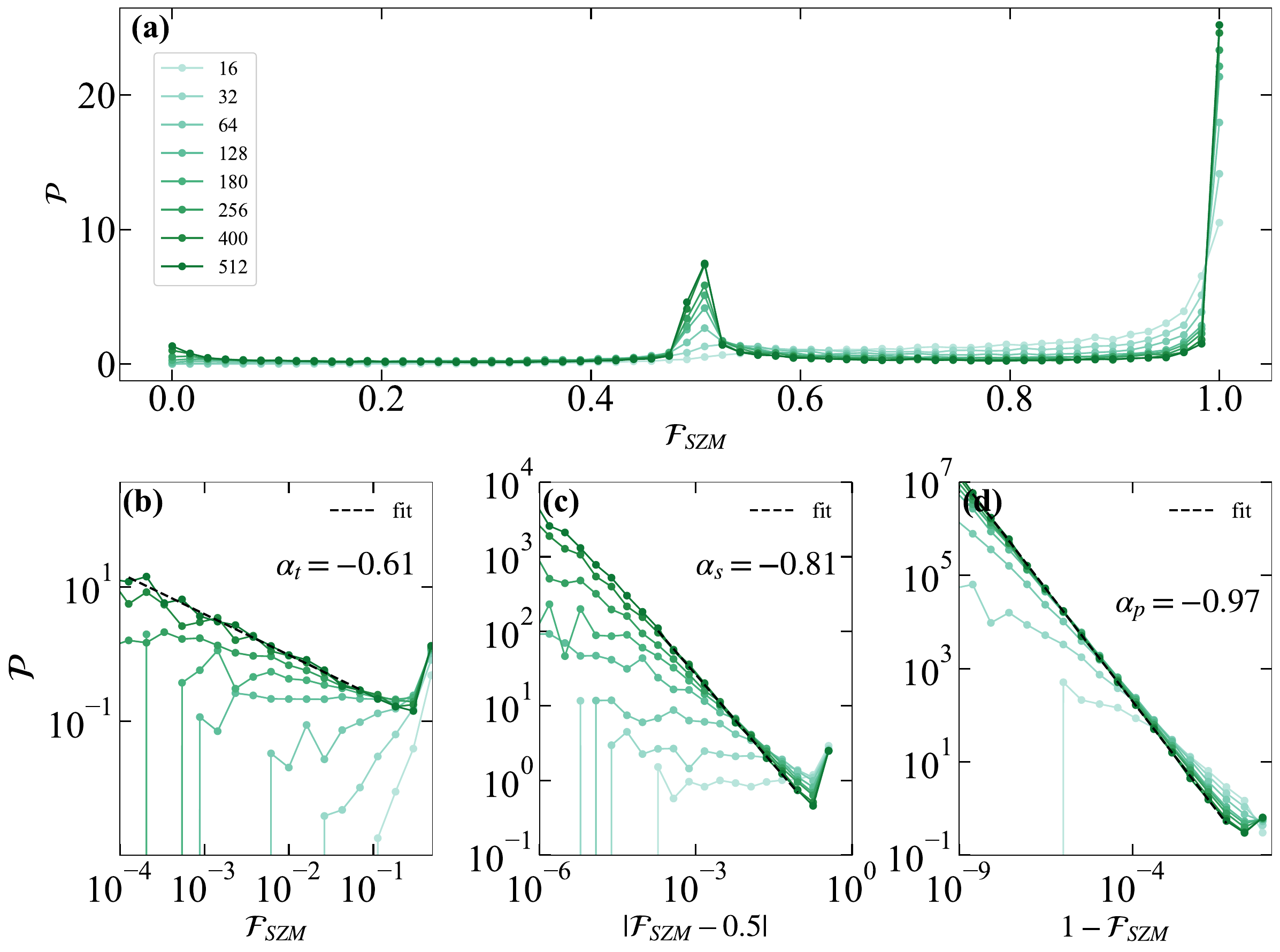}
    \caption{(a) Distribution of the symmetrized fidelity showing a primary peak at $1$ and a secondary peak at $0.5$ and a tertiary peak at $0$, (b)Logarithmic histogram of $\mathcal{F}_{\rm SZM}$ indicating a power law behaviour of the tertiary peak with an exponent $\alpha_t \approx -0.61$, (c) Logarithmic histogram of the $|\mathcal{F}_{\rm SZM}-0.5|$ showing a power law behaviour with exponent $\alpha_s\approx0.81$. (d) Logarithmic histogram of the $(1-\mathcal{F}_{\rm SZM})$ showing a power law behaviour with exponent $\alpha_p\approx0.97$ of the primary peak. The data in the plot correspond $2\times 10^4$ samples at disorder strength $W=3.0$ }
    \label{fig:fidelity_dist_ce} 
\end{figure}

\subsection{Fidelity in the trivial and the topological regimes}
A plot of the average and typical values of the symmetrized fidelity for the canonical ensemble is shown in figure \ref{fig:fidelity_full_phase_with_finite_size_corrections_can}. The qualitative features of the symmetrized fidelity in this case are similar to that of their counterparts in the microcanonical ensemble.\\
\textit{Trivial Regime}: Deep in the trivial region, we find no difference in the average and the typical values of $\mathcal{F}_{\rm SZM}$ in comparison to the trivial regime of the microcanonical ensemble. In both ensembles, the average value of fidelity shows a power law decay to zero with system size with a power close to $-1$, whereas the typical value shows an exponential decay to $0$.\\ 
%In contrast to the topological region, the finite size corrections in the trivial regime are the same in the two ensembles\\

\textit{Topological Regime}: In the topological regime, away from the critical point, both the average and the typical values of symmetrized fidelity approach $1$ with exponential corrections in the system size. In comparison to the topological regime of the  microcanonical ensemble, the decay length of the exponential is larger, both for the average and the typical values.
We also calculate critical exponents close to IRFP on the topological side of the transition using one parameter scaling, see fig \ref{fig:critical_exponents_canonical_topo}. Additionally we verify our calculation of critical exponents $\nu$ for the average and typical values of fidelity by minimizing a cost function $C(\nu)$. 
We find the critical exponents for both the average and the typical value of fidelity in the canonical case to be close to $\nu=2$, same as in the microcanonical case.

\subsection{SZM Fidelity at IRFP}
The most interesting dependence of the fidelity on the ensemble is at the infinite randomness fixed point. We find that the average and typical values $\overline{F_{\rm SZM}}$ and $\mathcal{F}_{\rm SZM}^{typ}$ saturate to values lower than their counterparts in the microcanonical ensemble. The decline in these values at IRFP for the canonical ensemble can be understood by studying their distributions over the ensemble.

\subsubsection{Distribution of Fidelities}
In figure \ref{fig:scaling_inv_fidelity_irfp_log_log_can}, we show the distribution of $\mathcal{F}_l$ and $\mathcal{F}_r$ at the IRFP for a canonical ensemble. The qualitative features of the distributions in the figure are similar to that in the microcanonical ensemble. However, a marked difference is in the height of the peak at $\mathcal{F}_{l/r}=0$ which is significantly more in the case of canonical ensemble. The peaks at $\mathcal{F}_{l/3}=\{0,1\}$ both show a power law behaviour to give a distribution of the form
\begin{equation}
\mathcal{P}\!\bigl(\mathcal{F}_{l/r}\bigr)=
  \begin{cases}
    \displaystyle\frac{W_p}{\bigl(1-\mathcal{F}_{l/r}\bigr)^{\alpha}} 
      & \text{if } \mathcal{F}_{l/r}\lesssim 1,\\[6pt]
    \displaystyle\frac{W_s}{\bigl(\mathcal{F}_{l/r})^{\beta}} 
      & \text{if } \mathcal{F}_{l/r}\gtrsim 0
      % \\[6pt]
      %     \displaystyle 0 
      % & \text{if } \mathcal{F}_{\mathrm{SZM}}< 0.5.
  \end{cases}
\end{equation}
for constants $W_p,W_S$.
Our results show the values of $\alpha$ and $\beta$ to be close to $1$, see figure \ref{fig:scaling_inv_fidelity_irfp_log_log_can}.

The distribution of $\mathcal{F}_{\rm SZM}$ for the canonical ensemble is shown in figure \ref{fig:fidelity_dist_ce}. Unlike in the case of microcanonical ensemble, here we get a triple-peak distribution fot the symmetrized fidelity. The three peaks are at $\mathcal{F}_{\rm SZM} = \{0,1/2,1\}$. The tertiary peak at $0$, which is absent in microcanonical ensemble is an indication that at the IRFP, there is a significant number of samples belonging to the canonical ensemble that have no remnants of topology. Instead, they are indistinguishable from the Anderson localized wavefunctions on the trivial side of criticality.\\ 

All the three peaks in the distribution of $\mathcal{F}_{\rm SZM}$ in the canonical ensemble demonstrate power-law behaviour, giving the distribution a form
\begin{equation}
\mathcal{P}\!\bigl(\mathcal{F}_{\mathrm{SZM}}\bigr)=
  \begin{cases}
    \displaystyle\frac{a_p}{\bigl(1-\mathcal{F}_{\mathrm{SZM}}\bigr)^{\alpha_p}} 
      & \text{if } \mathcal{F}_{\mathrm{SZM}}\lesssim 1,\\[6pt]
    \displaystyle\frac{a_s}{\bigl|\mathcal{F}_{\mathrm{SZM}}-0.5\bigr|^{\alpha_{s}}} 
      & \text{if } \mathcal{F}_{\mathrm{SZM}}\approx 0.5
      \\    \displaystyle\frac{a_t}{\bigl(\mathcal{F}_{\mathrm{SZM}})^{\alpha_{t}}} 
      & \text{if } \mathcal{F}_{\mathrm{SZM}}\gtrsim 0
      
      % \\[6pt]
      %     \displaystyle 0 
      % & \text{if } \mathcal{F}_{\mathrm{SZM}}< 0.5.
  \end{cases}
\end{equation}

From our numerical calculation, we get the values $\alpha_p = 0.97$, $\alpha_s = 0.81$ and $\alpha_t=0.61$.

% \newpage
\section{Setting up the microcanonical ensemble}\label{appendix:mcan_ensemble}
The target is to draw i.i.d samples from uniform distributions separately for $\{X_i\}$ and $\{h_i\}$ such that the average values of $\ln X$ and $\ln h$ are fixed for each sample. Below, we show the necessary steps required for the bond variables $\{X_i\}$; sampling the field variables $\{h_i\}$ can be done in the same way.
For a given system size $L$, with a disorder strength $W$ for each set of the bond variables $\{X_i\}$, $L-1$ variables $u_i$ are sampled from a uniform distribution $U(0,W)$ of mean $0$ and width $W$ such that $u_i\in[-W/2,W/2]$. Then a constant $c = \min \{u_i\} + \epsilon$ with a small tolerance value $\epsilon >0$ is added to each $u_i$ such that 
\begin{equation}
    \sum_{i=1}^{L-1}\ln(u_i+c) = (L-1)\overline{\beta}
\end{equation}
where $\overline{\beta}$ is the target mean of $\ln X$.
A function $f(c)$ is then defined as follows
\begin{equation}
    f(c) = \sum_{i=1}^{L-1}\ln(u_i+c) - (L-1)\overline{\beta}
    \label{eq:def_f(c)}
\end{equation}
Since $f(c)$ is a monotonically increasing function of $c$, there exists a $c^*$ such that 
$f(c^*)=0$. We then numerically solve for the $c^*$. The desired sample $\{X_i\}$ of size $L-1$ and a target mean $\overline{\ln X}=\overline{\beta}$ is then 
\begin{equation}
    \{X_i\} = \{u_i\} + c^*
\end{equation}

\section{Cost function for Critical Exponents}\label{appendix:critical_exponent_details}
Below we describe the cost function $C(\nu)$ and its calculation in order to obtain the critical exponents shown in figures \ref{fig:critical_exponent_topological_mCan} and \ref{fig:critical_exponents_canonical_topo}.
The description below is for the average value of fidelity $\overline{\mathcal{F}_{\rm SZM}}$. The calculation for the typical value is analogous.
We assume a single-parameter scaling form of $\overline{\mathcal{F}_{\rm SZM}}$ close to criticality. In the topological side of the critical point $\delta=0^+$,
\begin{align}
    1-\overline{\mathcal{F}_{\rm SZM}} = Y\Big(\frac{L}{\xi(\delta)}\Big)
\end{align}
and $\xi(\delta) = |\delta|^{-\nu}$.

The error associated with each $Y$ is the same as the error associated with $\overline{\mathcal{F}_{\rm SZM}}$, that is $s/\sqrt{N}$ for
\begin{align}
    s(L,\delta) = \sqrt{{\frac{1}{N-1}\sum_{j=1}^N}\Bigg(\mathcal{F}_{\rm SZM}^j(L,\delta) - \overline{\mathcal{F}_{\rm SZM}}(L,\delta)\Bigg)^2}
\end{align}
where $N$ is the number of disorder iterations.
 
Following this scaling form, for each $\delta$, we transform 
\begin{align}
    (L,Y(L,\delta))\longrightarrow (L|\delta|^{\nu}, Y)
\end{align}
If the scaling hypothesis holds, then all such curves (corresponding to different values of $\delta$) should collapse together. 

We divide the $X$-axis uniformly into $B$ bins.
For each bin, we define the inverse-error-weighted mean value of $Y_i$ in the mean and a cost function:
\begin{align}
    \mu_b = \frac{1}{n_b}\sum_i w_i Y_i
\end{align}
and 
\begin{align}
    c_b(\nu) =\sum_{i\in b}w_i(Y_i-\mu_b)^2
\end{align}
where $w_i = 1/s_i$ and $n_b$ is the number of points 
The cost function is then defined as 
\begin{align}
    C(\nu) = \frac{1}{B}\sum_{b} c_b(\nu)
\end{align}
which is minimized against $\nu$ to get the critical exponent.

\section{Calculating error bars using Bootstrap}
Bootstrap is a method to calculate the error in the estimation of the statistic of a parameter that does not require the knowledge of the distribution of that parameter. In the paper, we have used bootstrap to calculate the error in the scaling parameters $\xi$ in one-parameter scaling performed in figures \ref{fig:critical_exponent_topological_mCan} and \ref{fig:critical_exponents_canonical_topo}. Below we explain how the errors are calculated.
To begin with, we have a set of curves $Y(L;\delta)$ for several values of $\delta$. In figures \ref{fig:critical_exponent_topological_mCan} and \ref{fig:critical_exponents_canonical_topo} $Y=1-\mathcal{F}_{\rm SZM}^{typ}$ and $1-\overline{F_{\rm SZM}}$ respectively. In both the cases, we select as the reference curve the one corresponding to the largest magnitude of $\delta$. So $\delta_{ref} = \max \{\delta_i\}$ Let us denote this curve by $Y_{ref}(L)$. 
For the other curves, we perform a horizontal scaling $L\longrightarrow L/\xi$ in order to collapse them as closely as possible with the reference curve $Y_{ref}(L)$. The collapse is not perfect and $\xi$ is chosen by the method of least square error over the set of points $(L,Y(L,\delta))$. In this way $\xi(\delta)$ is calculated.
To obtain the error in $\xi$ is where bootstrap is implemented. 

In bootstrap method, an empirical distribution for $\xi(\delta)$ is obtained by repeatedly sampling the same datapoints in different possible ways.
Once we get an initial $\xi$ for a given curve, the points $L/\xi$ that lie within the domain of the reference curve is collected in a set $A$. Let us say there are $n$ such points. In each bootstrap round, $n$ points are picked from this set, each draw is followed by replacement of the point in the set. After $n$ draws 
\begin{align}
    S(\xi) = \frac{1}{n}\sum_{j\in A} \big(\log Y_{ref}(L)-\log Y(L/\xi)\big)^2
\end{align}
is calculated.
$\xi^*$ is the value of $\xi$ that minimizes $S(\xi)$. 
This procedure is repeated $N_B$ times. Finally, the error in $\xi$ is given by
\begin{align}
    \Delta\xi = \sqrt{\frac{1}{N-1}\sum_{j=1}^{N_b}\Big(\xi^*_j - \overline{\xi^*}\Big)^2}
\end{align}
where 
\begin{align}
    \overline{\xi^*} = \frac{1}{N_B}\sum_{j=1}^{N_B}\xi^*_j
\end{align}

\subsection{Color Gradient Scheme for Figure \ref{fig:fidelity_weight_collapse_mce}(a)}
\label{app: color_gradient}
For a given $(L,W)$, we split the distribution of $\mathcal{F}_{SZM}$ into two sets:
\begin{align}
    &f_>(L,W) = \{\mathcal{F}_{SZM}(L,W): \mathcal{F}_{SZM}(L,W)\geq \sqrt{8}/\pi\}\\
    & f_<(L,W) =  \{\mathcal{F}_{SZM}(L,W): \mathcal{F}_{SZM}(L,W)< \sqrt{8}/\pi\}
\end{align}
We plot the disorder-averaged values $\overline{f_>}$ and $\overline{f_<}$ against $W$ for different system sizes $L$.
To set up the color scale, we define
\begin{align}
    & s_>(L,W) = \frac{1}{N} [|\{\mathcal{F}_{SZM}(L,W): \mathcal{F}_{SZM}(L,W)\in [1-\epsilon, 1]\}|]\\
    & s_<(L,W) = \frac{1}{N} [|\{\mathcal{F}_{SZM}(L,W): \mathcal{F}_{SZM}(L,W)\in [0.5,0.5+\epsilon]\}|]
\end{align}
for a small number $\epsilon$ that we choose to be $0.05$.
Here $|\{\bullet\}|$ represents the cardinal number of the set and $N=2\times 10^4$ is the total number of samples realized for each $(L,W)$.
We normalize $s_>$ and $s_<$ as follows:
\begin{align}
    & \tilde{s}_>(L,W) = \frac{s_>(L,W)}{\max_{W'}(s_>(L,W'), s_<(L,W'))}\\
    & \tilde{s}_<(L,W) = \frac{s_<(L,W)}{\max_{W'}(s_>(L,W'), s_<(L,W'))}
\end{align}
The color intensity of the points $f_>(L,W)$ and $f_<(L,W)$ are proportional respectively to    $\tilde{s}_>(L,W)$ and $ \tilde{s}_<(L,W)$.

\twocolumngrid
\newpage
\bibliography{szm_ref}
\end{document}